\documentclass[sigconf]{acmart}

\usepackage{subcaption} %
\usepackage{makecell} %
\usepackage{acmart-taps}
\makeatletter
  \newcommand\EatSpacesHack{\@bsphack\@esphack}
\makeatother

  \renewcommand{\comment}[1]{\EatSpacesHack}
  \newcommand{\PostSubmission}[1]{\EatSpacesHack}
  \newcommand{\todo}[1]{\EatSpacesHack}
  \newcommand{\basi}[1]{\EatSpacesHack}
  \newcommand{\ak}[1]{\EatSpacesHack}
  \newcommand{\reviewfix}[1]{\EatSpacesHack}

  \usepackage[suppress]{color-edits}
  \addauthor{ak}{blue}

\newcommand{\V}[1]{\mbox{\textit{#1}}}  %

\def\Snospace~{\S{}}

\AtBeginDocument{%
  }

\begin{document}

\acmYear{2024}\copyrightyear{2024}
\setcopyright{rightsretained}
\acmConference[ACM FAccT '24]{ACM Conference on Fairness, Accountability, and Transparency}{June 3--6, 2024}{Rio de Janeiro, Brazil}
\acmBooktitle{ACM Conference on Fairness, Accountability, and Transparency (ACM FAccT '24), June 3--6, 2024, Rio de Janeiro, Brazil}
\acmDOI{10.1145/3630106.3659041}
\acmISBN{979-8-4007-0450-5/24/06}

\title[Auditing for Racial Discrimination in the Delivery of Education Ads]{Auditing for Racial Discrimination in the Delivery of Education Ads}
\author{Basileal Imana}
\email{imana@princeton.edu}
\orcid{0000-0002-6645-7850}
\affiliation{%
  \institution{Center for Information Technology Policy, Princeton University}
  \city{Princeton}
  \state{New Jersey}
  \country{USA}
  \postcode{08544}
}

\author{Aleksandra Korolova}
\email{korolova@princeton.edu}
\orcid{0000-0001-8237-9058}
\affiliation{%
 \institution{Department of Computer Science and School of Public and International Affairs, Princeton University}
  \city{Princeton}
  \state{New Jersey}
  \country{USA}
  \postcode{08544}
}

\author{John Heidemann}
\email{johnh@isi.edu}
\orcid{0000-0002-1225-7562}
\affiliation{%
 \institution{Information Sciences Institute, University of Southern California}
 \city{Los Angeles}
 \state{California}
 \country{USA}
 \postcode{90292}
}

\renewcommand{\shortauthors}{Imana et al.}

\begin{abstract}
Digital ads on social-media platforms play an important
  role in shaping access to economic opportunities.
Our work proposes and implements a new third-party
  auditing method that can evaluate racial bias in the \emph{delivery of ads for education opportunities}.
Third-party auditing is important because it allows external parties
  to demonstrate presence or absence of bias in social-media algorithms.
Education is a domain
  with legal protections against discrimination and concerns of racial-targeting,
  but bias induced by ad delivery algorithms has not been previously explored in 
  this domain.
Prior audits demonstrated
  discrimination in platforms'
  delivery of ads to users for housing and employment ads.
These audit findings supported legal action that
  prompted Meta to change their ad-delivery algorithms
  to reduce bias,
  but only in the domains of
  housing, employment, and credit.
In this work, we propose a new methodology that allows us to measure racial discrimination in a platform's
  ad delivery algorithms for education ads.
We apply our method to Meta using ads for real schools
  and observe the results of delivery.
We find evidence of 
  racial discrimination in Meta's algorithmic delivery of ads for education opportunities,
  posing legal and ethical concerns.
Our results extend evidence of algorithmic discrimination 
  to the education domain,
  showing that current
  bias mitigation mechanisms are narrow in scope,
  and suggesting a broader role for third-party auditing
  of social media in areas where ensuring non-discrimination is important.
\end{abstract}

\ccsdesc[500]{Social and professional topics~Technology audits}
\ccsdesc[500]{Social and professional topics~Systems analysis and design}
\ccsdesc[500]{Social and professional topics~Socio-technical systems}

\keywords{algorithmic auditing, targeted advertising, ad delivery, education ads, racial discrimination}

\maketitle

\section{Introduction}

Social-media platforms are a key method of advertising, and with the rapid growth of personalized advertising enabled by them, the platforms 
  play a significant role in shaping access to opportunities.
This role has been subject to increased scrutiny 
  through academic research~\cite{Ali2019a, Imana2021, Levi2022, Lam2023, Ali2023problematic},
  civil rights audits~\cite{FacebookCivilAuditProgress},
  and regulation in the U.S.~and E.U.~\cite{Nonnecke2022}.

Evidence of discrimination
  in how social-media platforms shape targeting and delivery of ads to users has been growing in recent years.
Initial reports showed that ad \emph{targeting} options,
  such as demographic attributes,
  could be used to discriminate~\cite{Speicher2018, ProPublicaGender, ProPublicaRace},
  leading platforms to limit the targeting options they make available for
  housing, employment and credit (HEC) ads~\cite{FbChanges2020a, GoogleChanges2020},
  each an economic opportunity with prior government oversight.
While removal of options for demographic targeting prevents directed discrimination,
subsequent audits showed discrimination can occur at ad \emph{delivery} stage,
  the algorithmic process by which a platforms
  decide when and which ads to display to users, and how much to charge the advertisers for them.
  Specifically, Meta's ad delivery algorithms were shown to be discriminatory by race and gender in the
 delivery of housing and employment ads,
  even when an advertiser targeted
  a demographically balanced audience~\cite{Ali2019a, Imana2021, Levi2022}.
Follow-up studies showed ad delivery algorithms
  play an active role in shaping access to information on politics~\cite{Ali2019b}
  and climate change~\cite{Sankaranarayanan2023} in a way that is not
  transparent to both advertisers and end-users.

In response to these findings and as part of a legal settlement with the
  U.S.~Department of Justice, in 2023
  Meta deployed a Variance Reduction System (VRS)
  to reduce bias in delivery of housing ads
  ~\cite{FacebookvsHUD2, FacebookvsHUD2023TR},
  and subsequently for employment and credit ads~\cite{FacebookvsHUD2023}.
VRS's goal is to ensure the fraction of an HEC ad's impressions allocated to users of a particular gender (resp. race) does not deviate too much from the fraction of impressions allocated to users of that gender (resp. race) among all ads shown to them over the last 30 days~\cite{FacebookvsHUD2}. %
Meta achieves this goal by making adjustments to the bidding strategy Meta executes on behalf of the advertiser.

Although the progress towards de-biasing delivery of
   HEC ads is promising, concerns remain that
   delivery of other types of opportunities advertised on social-media platforms may be discriminatory.
In particular, in the U.S., the potential for ad delivery discrimination is 
  a concern in domains such as
  insurance, education, healthcare, or
  other ``public accommodations'',
  a legal term that encompasses various types of
  businesses that provide service to the general public~\cite{EducationAct, Lerman1978}.
No methodology is known to audit ad delivery in those domains.
Developing new methods can inform
  applications of existing anti-discrimination law in these domains,
  similar to how prior audits have influenced changes to the systems and informed guidance on applications of Fair Housing Act
  to digital platforms~\cite{FHAGuidance}.

\textbf{Challenges for Auditing Ad Delivery:}
Auditing for discrimination in ad delivery is challenging for three reasons.
First, the algorithms that select ads to be displayed
  and choose the prices to charge for them are opaque
  to advertisers and users.
The \emph{ad delivery algorithms} used for these choices are
  a closely guarded ``secret sauce'' of each platform, as they are central to platform monetization.
These algorithms often use machine-learning-based predictions to quantify how
  ``relevant'' an ad is to the
  user~\cite{FacebookAdAuction, TwitterAdAuction, LinkedInAdAuction} and
  how likely a user is to be valuable to the advertiser~\cite{FacebookvsHUD2023TR}.
Furthermore, the algorithms often decide on the advertiser's
  behalf how much money to spend on trying to get their
  ad in front of the particular user~\cite{FacebookvsHUD2023TR, Lambrecht2016}. 
The concern is that these algorithms, 
  designed to achieve the platform's long-term business goals,
  may propagate or exacerbate historical biases,
  or may learn biases from training data.

A second challenge is the limited access
  auditors have to data crucial to examine
  how platforms' algorithms work and what results they produce.
The state-of-the-art in auditing ad delivery
  algorithms takes a black-box approach,
  utilizing only the limited features available to advertisers~\cite{Imana2021, Imana2023}.
These approaches rely on creating target ad audiences with a specific demographic composition,
  a challenging process that
  often requires auxiliary data sources to provide demographic attributes.
Additionally,
  in conventional evaluations for algorithmic fairness,
  the auditor has access to individual- or group-level data, including scores assigned by the algorithms to those individuals.
However, 
  black-box auditors of ad delivery algorithms
 have access only to
  aggregated reports from the platform about who is reached.
These reports distance the auditor from the evaluation
  through several levels of indirection, 
  since they do not identify demographics of interest such as race,
  nor do they provide data on the relevance or value scores assigned by the~platform's algorithms they are trying to evaluate.

Thirdly, the use of a black-box evaluation of platform algorithms
  must control for confounding factors %
  that may affect ad delivery, such as temporal effects and market forces.
For example,
  an observation of an ad's delivery to relatively larger number
  of White individuals than Black individuals
  does not necessarily imply discrimination
  in the ad delivery algorithm against Black users.
Instead,
  it could be attributed to higher number of White individuals
  using the platform during the observation period.
Another reason for differences in delivery could be
  competition with other advertisers 
  which have larger budgets for some racial groups~\cite{Lambrecht2016, DI2019}.
The work of Ali and Sapiezynski et al.~was the first to control for
  such factors by running \emph{paired ads} that are equally affected by the confounding factors,
  and evaluating \emph{relative difference} in the delivery
  of the ads for different demographic groups~\cite{Ali2019a}.
A subsequent study considered differences in job qualification as a relevant
  confounding factor for evaluating whether bias in delivery of employment ads constitutes discrimination,
  and proposed a method to control for it
  using a particular type of paired ads~\cite{Imana2021}.

\textbf{Why Education?}   
In this work, we explore \emph{education} as a new domain for which
  to study the potential role of ad delivery for discrimination.
Avoiding discrimination in the delivery of educational opportunities
 matters because
 education has long-term impact on the career
 and financial well-being of individuals~\cite{FederalReserve2023}.
One specific concern in U.S.~higher education is 
  the potential for for-profit colleges to produce
  poor financial outcomes for their graduates~\cite{Deming2013, Looney2015}.
Such colleges spend a large amount on advertising~\cite{Deming2013},
  and have been accused of disproportionately targeting racial minorities~\cite{Kahn2019}.
Prior work has not studied whether ad delivery algorithms
  propagate historical racial targeting biases in such colleges,
  a unique aspect of the education domain. 
In addition,
  education is an important category because of advertiser large spending in this domain.
According to Meta's whitepaper (\cite{wernerfelt2022estimating}, Figure 4(a)),
  education is one of the top eight verticals among its advertisers,
  significantly ahead of verticals
  such as political advertising which have received much scrutiny.

\textbf{Why Disparate Impact?} 
We specifically consider \emph{disparate impact} in education ad delivery
  because it is an outcome-based legal doctrine for assessing discrimination
  that is agnostic to the reasons for such outcome~\cite{Barocas2016, datta2018discrimination}.
Under this legal doctrine,
  disparate impact occurs 
  if the outcome of ad delivery differs significantly by a demographic
  attribute such as race.
The burden for understanding and justifying the source of the bias
  as legally permissible shifts to the platform.
We expand on the implications of our work for legal liability in \autoref{sec:legal_liability}.

\textbf{Our Contributions:}
Our first contribution is %
  a new method for testing for the presence of racial discrimination
  in how platforms deliver \emph{education ads} (\autoref{sec:education_method}).
Inspired by a methodology developed in prior work for
  testing for gender discrimination in job ad delivery~\cite{Imana2021},
  we extend this approach racial discrimination in education ads
  by identifying colleges with different historical recruitment by race.  
By pairing ads for a for-profit and a public college,
  then looking at relative differences in ad delivery by race,
  we can isolate the role of platform's algorithm
  from other confounding factors.
The insight in our method is to employ a pair of education ads for seemingly equivalent opportunities to users, but with one of the opportunities, in fact, tied to a historical racial disparity that ad delivery algorithms may propagate.
If the ad for the for-profit college is shown to relatively more Black users than the ad for the public college, we can conclude that the algorithmic choices of the platform are racially discriminatory.

Our second contribution is to apply our method to Meta, where
  \emph{we detect racial discrimination in the delivery of education ads} (\autoref{sec:bias_result}).
We first evaluate the platform's algorithm using neutral ad creatives
  that control for confounding factors.
Our experiments with neutral creatives demonstrate that
 Meta's
  ad delivery algorithm shows ads for for-profit
  schools to relatively more Black users, and the difference is statistically significant for 
     two out of three pairs of schools we study.
In additional experiments,
  we show that when we use
  realistic ad creatives that the schools use in practice,
  the racial skew in delivery is \emph{increased} (\autoref{sec:realistc_ads}).
These results provide strong evidence
  that
  Meta's algorithms 
  shape the racial demographics of who sees which education opportunities,
  providing new evidence of Meta's potential
  long-term impact on careers and the financial well-being
  of individuals.
Our results also open questions of Meta's legal
  liability under the doctrine of disparate impact 
  of discrimination~\cite{datta2018discrimination, Barocas2016}.

Finally,
  we use our methodology to investigate whether Meta's algorithm
  steers delivery of ads disproportionately by race for
  schools whose enrollment or marketing practices have previously received
  legal scrutiny.
Over-delivery of ads for such schools
  may inadvertently disadvantage students who are not aware of the potential risks,
  such as poor labor market outcomes~\cite{Deming2013}.
We pick ads for three such schools 
  from a list compiled by Veterans Education Success in 2018~\cite{DOESuedSchools}
  and evaluate the
  racial difference in delivery of their ads compared to delivery of ads for
  public schools.
We find that Meta shows the ads for the selected schools
  to relatively more Black individuals.
Our results show that it may not be sufficient for these schools to target
  their ads equitably, but that ad platforms also need to ensure
  their algorithms are not introducing biases along legally protected characteristics
  such as race.
  
Our findings of discrimination in delivery of ads for real education opportunities
  show the
  issue of algorithmic discrimination is not limited to housing,
  employment, and credit opportunities and raises a broader question of whether
  the narrow scope of current solutions proposed by platforms
  such as Meta's VRS are sufficient.
Our work thus underscores the need for platforms,
auditors and regulators to expand the set of domains where impacts of ad delivery algorithms are considered.  

\section{Methodology: Using Pair of Schools with De-Facto Racial Skew in Enrollment}
    \label{sec:education_method}
We now describe how we test for discriminatory ad delivery
  in the education domain
  by pairing ads for for-profit and public schools
  while controlling for confounding factors.
  
The key insight of our design is to identify
  a historical racial disparity in the higher-education sector that we hypothesize may be
  propagated by ad delivery algorithms.
Specifically, we use known
  differences in enrollment of Blacks
  among for-profit and public schools
  to select content and landing pages for a pair of ads
   that are designed to
  probe the platform's algorithms
  and test whether it perpetuates the existing differences,
  even when both ads and their descriptions appear as
  similar education opportunities to users.
Our approach is inspired by prior research that
  used a pair of job opportunities requiring identical qualifications
  but exhibiting a de-facto gender skew among different employers~\cite{Imana2021}.
 
\subsection{Identifying a De-facto Skew In Education}
  \label{sec:defacto_skew_education}

We identify candidate schools to advertise for
  based on the de-facto racial skew towards Black students in
  for-profit colleges and towards White students in public colleges.
Per 2022 College Scorecard data from
  the U.S. Department of Education~\cite{CollegeScorecard},   
  Black students make up 25\% of the student body
  at for-profit colleges,
  whereas they account for only 14\% of students
  at public colleges.
This difference serves as the basis for designing our experiments
  to assess whether the ad delivery algorithms
  lead to discriminatory outcomes.
From the list of all public and for-profit colleges in
  College Scorecard data,
  we first build two shortlists of four-year colleges
  that have a de-facto racial skew,
  one for for-profit schools whose demographics
  skews towards Black students,
  and the second for public schools whose demographics skews towards
  White students.
Our hypothesis is that,
  if a platform's algorithm for education ad delivery is discriminatory
  in a way consistent with the de-facto skew,
  a for-profit school ad will
  be delivered to a disproportionately larger fraction of Black users than the public school ad.

Our hypothesis for the potential for this methodology to showcase discriminatory delivery
  stems from knowing platforms' ad delivery algorithms
  factor in historical data.
The algorithms are trained on
  data about relationships between users and entities they interact with
  collected from myriad of sources on and off the platform~\cite{Gerlitz2013},
  and from both online and offline sources~\cite{Hitlin2019}.
Historically, for-profit colleges have disproportionately 
  targeted racial minorities~\cite{Kahn2019},
  which also reflects in the current demographics of the students in those schools.
We know platforms' ad delivery algorithms
  consider not only a particular user's
  prior interactions with colleges, 
  but also interactions of other ``similar'' users~\cite{MetaPatentAds2011}.
Therefore, a Black person may receive a higher relevance score
  for a for-profit college ad than a White person because,
  historically, other Black people have interacted with the
  school or been targeted by similar schools.
Our method is designed to take advantage of this kind of ``learned'' bias to interrogate the algorithm for disparate outcome discrimination.

\subsection{Identifying a Pair of Education Opportunities to Minimize Confounding Factors}
  \label{sec:pair_ads_education}
  
Based on the de-facto enrollment skew,
  we next identify pairs of schools
  that offer educational opportunities
  that are equivalent in terms of selectivity and availability of online programs,
  but differ in their for-profit status.
We narrow down our shortlist of for-profit and public schools to
  those that are not very selective
  (acceptance rate $\approx > 50\%$) to
  minimize potential skew in delivery due to
  differences in educational qualifications
  among users in our audience.
We select schools that offer online, part-time degree
 programs
 to minimize effects of school location relative to 
  the location of our target audience.

Additionally, we aim to ensure the platform has sufficient signal and data
  about the schools we pick.
We thus further narrow down the list to schools that have
  at least 5,000 students,
  have an active page on the Meta platform,
  and actively run ad campaigns on the platform.

\aptLtoX[graphic=no,type=html]{\begin{table}
    \centering
    \begin{tabular}{ p{3.7em}|p{10em}|p{11em} }
Pair ID & For-profit School & Public School  \\ 
\hline 
epair-1a & Strayer University \newline (B=79\%, W=13\%, O=8\%) \newline (Admit: 100\%) & Colorado State University \newline (B=7\%, W=64\%, O=29\%) \newline (Admit: 98\%)   \\ %
\hline 
epair-2a & American InterContinental University \newline  (B=29\%, W=26\%, O=45\%) \newline (Admit: 100\%) &  Fort Hays State University \newline  (B=2\%, W=50\%, O=48\%) \newline (Admit: 91\%)  \\  %
\hline 
epair-3a & Monroe College \newline  (B=42\%, W=3\%, O=55\%) \newline (Admit: 49\%) &  Arizona State University \newline  (B=7\%, W=58\%, O=35\%) \newline (Admit: 73\%)  \\  %
\end{tabular}
      \caption{List of pairs of schools we use in our experiments on delivery of education ads. For each school, the table shows the racial makeup of the student body (``B'' = Black students, ``W'' = White students, ``O'' = Other) and the admission rate.}
      \label{tab:education_ad_list}
\end{table}
\begin{figure}
    \centering
    \includegraphics[width=\columnwidth]{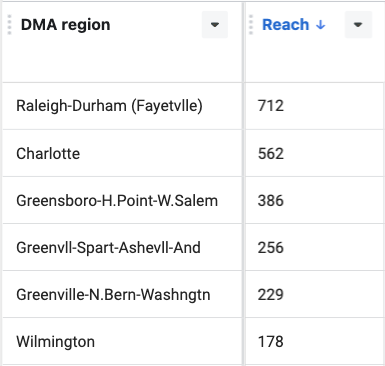}
    \caption{A partial screenshot of Meta's Ads Manager that demonstrates aggregate location data that Meta reports for ad recipients.}
    \label{fig:agg_location}
\end{figure}}{\begin{figure*}[t]
  \begin{minipage}{0.69\textwidth}
    \centering
    \begin{tabular}{ p{3.7em}|p{10em}|p{11em} }
Pair ID & For-profit School & Public School  \\ 
\hline 
epair-1a & Strayer University \newline (B=79\%, W=13\%, O=8\%) \newline (Admit: 100\%) & Colorado State University \newline (B=7\%, W=64\%, O=29\%) \newline (Admit: 98\%)   \\ %
\hline 
epair-2a & American InterContinental University \newline  (B=29\%, W=26\%, O=45\%) \newline (Admit: 100\%) &  Fort Hays State University \newline  (B=2\%, W=50\%, O=48\%) \newline (Admit: 91\%)  \\  %
\hline 
epair-3a & Monroe College \newline  (B=42\%, W=3\%, O=55\%) \newline (Admit: 49\%) &  Arizona State University \newline  (B=7\%, W=58\%, O=35\%) \newline (Admit: 73\%)  \\  %
\end{tabular}
      \captionof{table}{List of pairs of schools we use in our experiments on delivery of education ads. For each school, the table shows the racial makeup of the student body (``B'' = Black students, ``W'' = White students, ``O'' = Other) and the admission rate.}
      \label{tab:education_ad_list}
    \end{minipage}%
      \hfill
    \begin{minipage}{0.27\textwidth}
    \centering
    \includegraphics[width=\columnwidth]{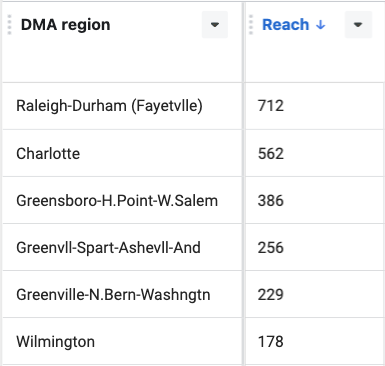}
    \captionof{figure}{A partial screenshot of Meta's Ads Manager that demonstrates aggregate location data that Meta reports for ad recipients.}
    \label{fig:agg_location}
  \end{minipage}
\end{figure*}}
  
Using the above criteria,
  we narrow down our shortlist to 9 public schools
  and 3 for-profit schools.
We sort the for-profit schools based on the descending order
  of the differences between the percentages of Black and White
  students enrolled.
Simultaneously, we sort the public schools in descending order of the
  differences between the percentages of White and Black students enrolled.
We then group the three for-profit schools with a
  public school from the corresponding location in the sorted lists.
The pairs of schools selected for our experiments and the racial makeup of each school's student body are shown in Table~\ref{tab:education_ad_list}.
Although the schools in each pair
  may be differently ranked academically,
  we do not think this difference affects our evaluation
  of discrimination.
For both types of schools, non-discrimination should result in
  similar ad delivery to the different racial groups regardless of school ranking. 

We pick schools that are correlated with race in terms of school
  demographics and for-profit status to elicit bias that may exist
  in the ad delivery algorithm.
Picking schools in such a way may introduce
  confounding factors
  such as differences in familial or social ties to alumni~\cite{Hurwitz2011}
  and familiarity with a school's brand~\cite{Joseph2012};
  these may affect how likely a prospective student is to be admitted or
  how relevant the opportunity is to the student.
Consideration of ties with alumni, in particular, is a controversial practice
  among elite schools that
  is known to affect the
  racial composition of admitted students~\cite{Arcidiacono2022},
  and is being challenged by the U.S.~Department of Education~\cite{DOEvsHarvard}.
Our approach minimizes the effect of such factors by picking non-elite schools and targeting
  a large sample of random individuals from
  U.S.~states
  that are not tied to the specific location of the schools
  (see \autoref{sec:audience_education} and \autoref{sec:state_proxy_results} for details).
Furthermore, because platforms currently
  do not provide special access to support auditing,
  it limits the conceivable confounding
  factors we can control for.

\subsection{Running Ads on Meta and Evaluating their Delivery}
  \label{sec:how_we_run_ads}
Having described the key insight of our methodology
  based on which we identify pairs of education opportunities to advertise for,
  we next describe the steps we take
  to actually run the ads on Meta's ad platform.
  
At a high-level, the steps include building an ad audience,
  selecting ad creatives, budgets and other ad campaign parameters,
  launching the ad campaigns, collecting data on their delivery, and then using that data to evaluate skew in how the ads are delivered by race.
Meta does not allow targeting by race and does not return information about the race of ad recipients, so our ad audience construction and evaluations of
the ad's performance are crafted to be able to make such inferences.

\subsubsection{Building Audiences so that Location Reports Correspond to Race}
  \label{sec:audience_education}

\begin{table*}
\caption{List of voter datasets we use to construct ad audiences.}
\label{tab:voter_data_list}
\centering
\small
\begin{tabular}{ c|c|ccc }
\textbf{Group ID} & \textbf{Group of DMAs} & \makecell{\textbf{\# of Blacks}} & \makecell{\textbf{\# of Whites}} \\
\hline
Group 1 & Raleigh-Durham, Wilmington, Greenville-Spartaburg, Norfolk-Portsmouth & 697,492 & 2,282,243  \\ 
Group 2 & Charlotte, Greensboro,  Greenville-New Bern  & 818,599 & 2,564,627 \\ 
\end{tabular}
\end{table*}

Following the approach from prior work~\cite{Ali2019a}, we build the target audiences for our ads in a way that will allow us to infer
  the race of ad recipients from Meta's reports on locations of ad recipients.
We use two features of Meta's advertising system to do so -- its reports on the number of impressions an ad receives broken down by Designated Market Area (DMA);
  and the Custom Audience feature, that allows advertisers
  to create a target audience using a list of personally identifiable information,
  such as names, email and home addresses~\cite{FBCustomAud}. 
Figure~\ref{fig:agg_location}
  shows a screenshot of Meta's Ads Manager portal, illustrating a report provided by Meta to the advertiser, of the
  locations of the ad impressions aggregated
  by DMAs.

We construct ad audiences
  by (DMA, race) pairs,
  so that we can infer the breakdown of delivery by race from the breakdown of delivery by DMA provided by the platform. 
We rely on voter datasets
   that contain race information of individuals, and build each Custom Audience so that half of it consist of White people from one group of DMAs and another half of it consists of Black people from another group of DMAs (non-overlapping with the first group).
For our experiments, we use publicly available voter dataset from 
   North Carolina (NC)~\cite{NCVoterData} (see the summary statistics in Table~\ref{tab:voter_data_list}).
For example,
  say we include
  only Black individuals from Raleigh DMA and
  only White individuals from Charlotte DMA in our ad audience.
Then, whenever our ad is shown in Raleigh, we can infer it was shown
to a Black person and, when it is shown in Charlotte -- we can infer it was shown to a White person.  

If an ad is delivered to a user
  outside the DMAs listed in Table~\ref{tab:voter_data_list}, for example, due to people traveling,
  we disregard the impression in our evaluation
  since we cannot infer the race of the user.
To ensure location does not skew our results,
  we replicate all our experiments using
 ``flipped'' audiences where we
 reverse the group of DMAs from which we pick
 Black and White individuals we include in our audience.

To evaluate reproducibility of our auditing results
  without introducing test-retest bias,
  we repeat our experiments on randomly selected audience partitions
  that are subsets of the voter lists.
Each partition contains 15K White and 15K Black individuals,
  which we find by running test ads
  is a large enough audience size
  to get enough samples for our experiments.
We conducted our test ads on partitions distinct from those we use in our experiments
  to evaluate bias in ad delivery.
Like prior audits of ad delivery~\cite{Ali2019a, Imana2021},
  our methodology does not require targeting
  an equal number of Black and White users for validity, but
  we still do so for ethical reasons to avoid
  discriminating as part of conducting our audits.
We name each audience partition based on
  whether the audience is a flipped version or not.
For example, a partition named ``{aud-nc-1}''
  indicates we included Black individuals from DMA group 1 (from Table~\ref{tab:voter_data_list}) and
  White individuals from DMA group 2;
  and a partition named ``{aud-nc-1f}'' is a flipped
  version of the audience.

\subsubsection{Selecting Ad Creatives and Campaign Parameters}
  \label{sec:edu_ad_creatives}

For ad creatives,
  we first use ad text and images that are neutral and consistent
  across each pair of ads to control for the possible effects due to creative choices~\cite{Levi2022}.
We then run ads for the same schools using realistic
  creatives taken from each school's Meta Ad Library page
  to test how the delivery algorithm may
  amplify implicit cues in ad creatives.
  
\begin{figure*}
\centering
  \begin{minipage}{.5\textwidth}
  \centering
   \begin{subfigure}{0.49\columnwidth}
  \centering
  \includegraphics[width=\columnwidth]{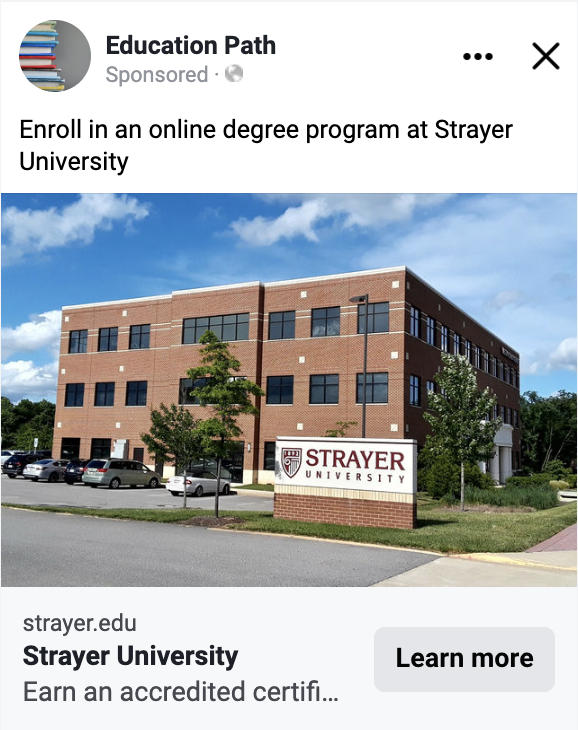}
  \caption{For-profit (neutral creative)}
  \label{fig:education_ads_for_neutral}
  \end{subfigure}
  \begin{subfigure}{0.49\columnwidth}
  \centering
  \includegraphics[width=\columnwidth]{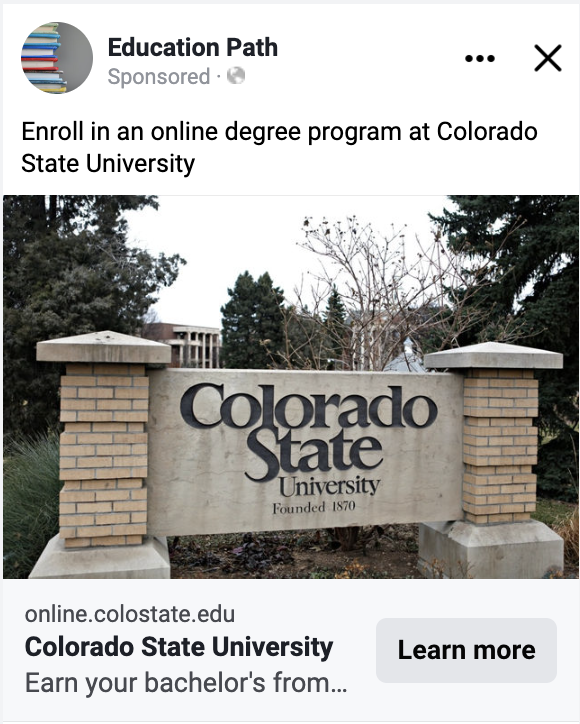}
  \caption{Public (neutral creative)}
   \label{fig:education_ads_pub_neutral}
  \end{subfigure}
\end{minipage}%
\begin{minipage}{.49\textwidth}
  \centering
  \begin{subfigure}{0.39\columnwidth}
  \includegraphics[width=\columnwidth]{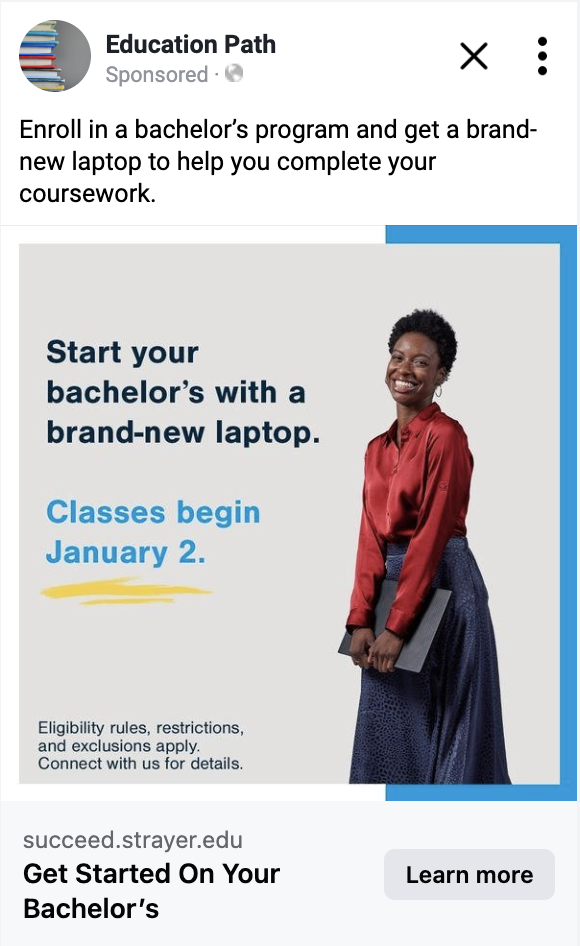}
  \caption{For-profit (realistic)}
   \label{fig:education_ads_for_realistic}
  \end{subfigure}
  \begin{subfigure}{0.40\columnwidth}
  \includegraphics[width=\columnwidth]{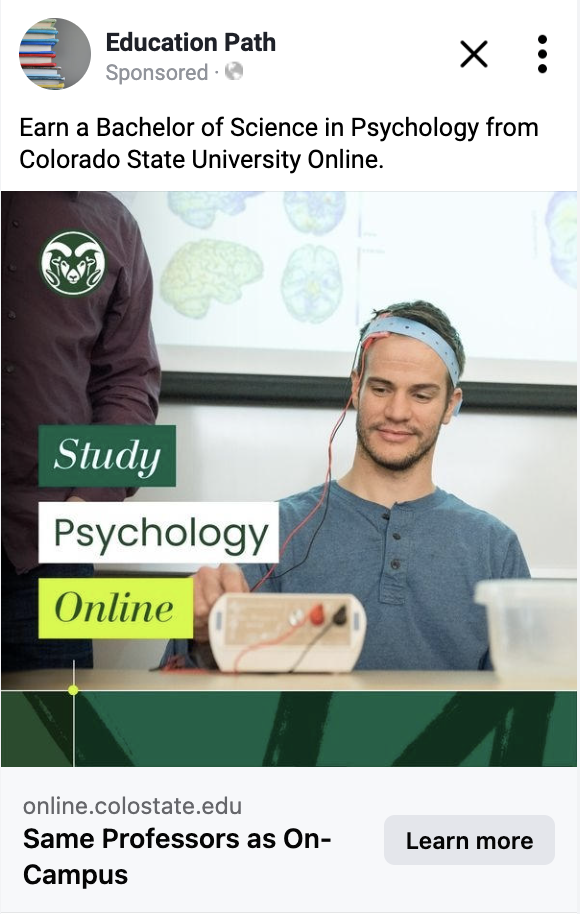}
  \caption{Public (realistic)}
   \label{fig:education_ads_pub_realistic}
  \end{subfigure}
\end{minipage}
  \caption{Example ad creatives for studying racial skew in delivery of education ads. The two left figures use neutral ad creatives that do not include people. The two right figures use realistic creatives taken from each school's ad library page and include people of a specific perceived race.}
  \label{fig:education_ads_both}
\end{figure*}

\emph{Neutral creatives:}
To minimize the possible influence of ad creative choice on delivery,
  we ensure the image and text of our ads
  are consistent and neutral.
For each school,
  we use a picture of the school's campus or logo,
  and avoid using images with people's faces,
  which prior
  work has shown influences ad delivery~\cite{Levi2022} or may influence users' engagement~\cite{Nagaraj2023}.
We use a consistent headline text for all ads that prompts recipients
  to enroll in an online degree program,
  with the sole difference being the incorporation of the respective school's name.
The destination sites, however, are different for each ad --
  they link to the school's official website dedicated
  to online programs.
We do so to ensure
  participants who are interested in the ad are provided access to the actual opportunity.
 Figure~\ref{fig:education_ads_for_neutral}  and Figure~\ref{fig:education_ads_pub_neutral}
  show an example pair of neutral
  ad creatives we use in our experiments.

\emph{Realistic creatives:}
To test the effects of ad delivery for
  real-world ads used by schools,
  we also run experiments using realistic ad creatives.
We take a snapshot of a list of ads run by the schools
  from Meta's public Ad Library~\cite{MetaAdLib}
  and manually annotate the images used in the ads by (perceived) race.
The manual effort  
     introduces the potential for annotation bias but extreme precision is not mandatory for our methodology.
For each school,
  we select from the list one representative ad that includes
  the face of an individual whose race
  is represented in majority of the ads.
 \autoref{fig:education_ads_for_realistic} and \autoref{fig:education_ads_pub_realistic} show an example pair of
  realistic ad creatives we use in our experiments.

Similar to prior work~\cite{Ali2019a, Imana2021},
  we run each pair of ads in an experiment simultaneously with the same
  campaign parameters including
  the budget, the audience, and
  time duration for the ads.
Running a pair of ads in such a way
  controls for temporal
  factors and market effects that may otherwise
  confound our measurement, and thus allows to isolate the role of ad delivery for any differences in outcomes.
We run all our ads with a ``Traffic'' objective
  that aims to increase traffic to the website that our
  ads link to~\cite{FBAdObjectives}.
We run them for a full 24 hours with a total
  budget of \$50 per ad.
We do not label our education ads as ``Special Ad Category'',
  as only housing, employment, credit and social issue ads are required to do so.
We limit targeting to the Custom Audiences
we built
  based on voter data and limit
  delivery to United States.
We do not add any
  additional targeting parameters.

\subsubsection{Launching and Monitoring Performance}
  \label{sec:monitor_perf}

We launch and monitor our ads using
  APIs that
  Meta provides to advertisers.
We do not launch our experiments until both ads
  have been approved by Meta.
Once our experiment starts and the pair of ads starts
  being shown to users,
  we use a script that fetches ad performance data
  at least once every hour to track delivery over time.
Because we build our audience in such a way that
  uniquely maps a location of a recipient to their race, %
  we use the DMA attribute that Meta provides
  to calculate the number of unique impressions by location,
  and in turn, by the corresponding race.

\subsubsection{Evaluating Skew}
  \label{sec:skew_eval}
  
We apply a skew metric to the racial
  breakdown of unique ad impressions to
  evaluate statistical significance of any racial skew
  we observe in the delivery of the pair of education ads.
We use a metric that is established in the literature
  for comparing the delivery of a pair of ads~\cite{Ali2019a, Ali2019b, Imana2021}.
We next introduce the notations and statistical test for the metric
  that underlie the empirical findings we
  present in~\autoref{sec:bias_result}.
  
Let $n_{f,b} $ and  $n_{f,w}$
  represent the number of users
  that saw the for-profit school ad, and are
  Black and White, respectively.
We observe these numbers from Meta's reported ``Reach'' metric on the ad's performance,
  which corresponds to unique impressions (number of people that saw the ad).
One can define the same terms for the public school ad ($n_{p,b}$ and $n_{p,w}$).
We can calculate the fractions of Black users
among the for profit school ad's and public school ad's recipients as:

\begin{equation*}\label{eq:true_fractions}
 s_{f,b} = \frac{{n_{f,b}}}{ n_{f,b} + n_{f,w}}  \hspace{0.5em}\text{ and }\hspace{0.5em}  s_{p,b} = \frac{{n_{p,b}}}{n_{p,b} + n_{p,w}}.   
\end{equation*}

We apply a statistical test to compare $s_{f,b}$ and $s_{p,b}$
  and evaluate whether there is statistically significant racial skew in ad delivery.
When the ad delivery algorithm is not skewed,
  we expect $s_{f,b} = s_{p,b}$
  because we ensure other confounding factors affect both ads equally.
As long as the two fractions are equal, even if they are not equal to $0.5$,
  there is no delivery bias.
For example, both ads may be delivered to 60\% Blacks
  due to fewer White people being online at the time of the experiments or other advertisers bidding higher for Whites than Blacks, and thus this skewed outcome would not be due to ad delivery algorithm's bias.
However, if there is a \emph{relative difference} between $s_{f,b}$ and $s_{p,b}$,
  we can attribute it to choices made by the platform's ad delivery algorithm.

\begin{figure*}
\centering
\begin{minipage}{.45\textwidth}
  \centering
  \begin{subfigure}{0.95\columnwidth}
    \includegraphics[width=1\columnwidth]{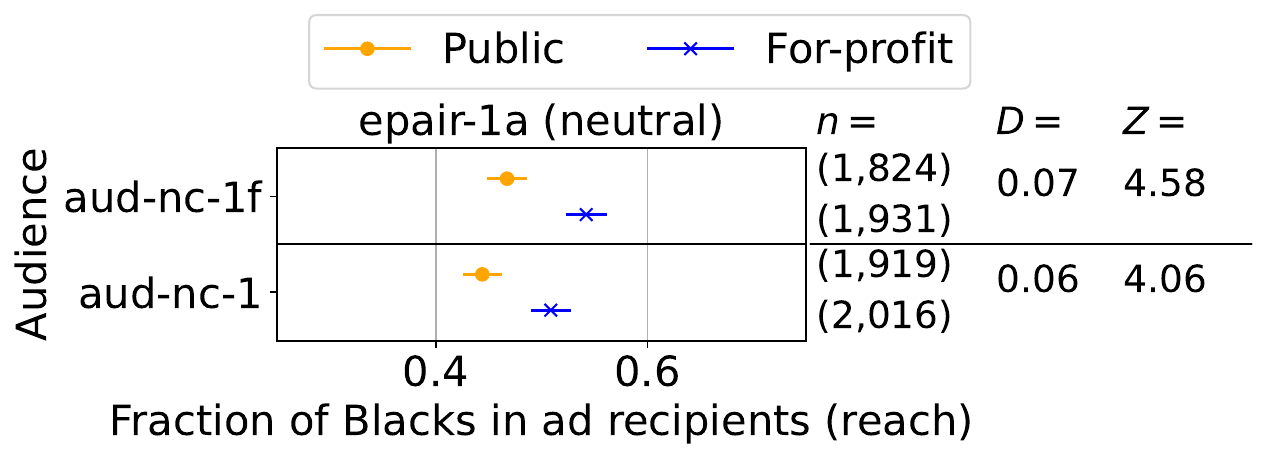}
    \caption{Strayer vs. Colorado State (neutral creatives)}
    \label{fig:epair-1-summary}
  \end{subfigure}
  \begin{subfigure}{0.95\columnwidth}
    \includegraphics[width=1\columnwidth]{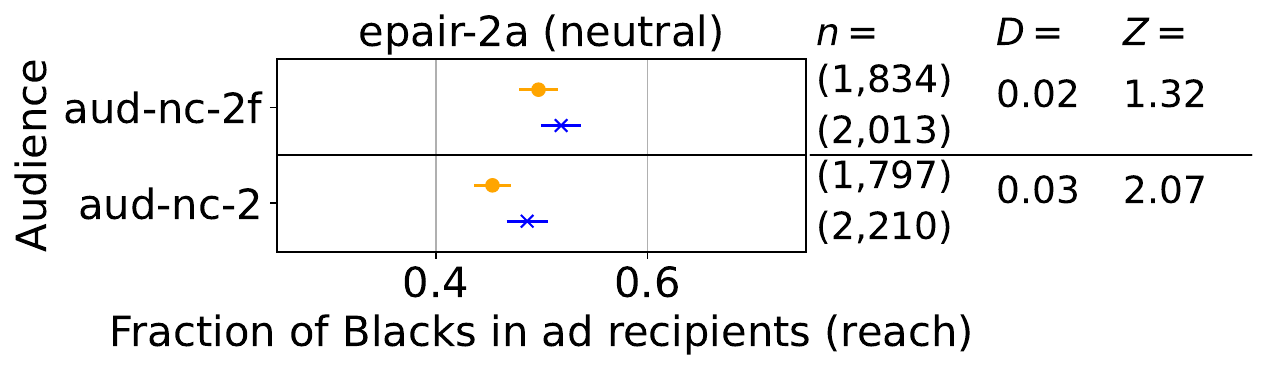}
    \caption{American Inter. vs. Fort Hays (neutral creatives)}
    \label{fig:epair-2-summary}
  \end{subfigure}
  \begin{subfigure}{0.95\columnwidth}
    \includegraphics[width=1\columnwidth]{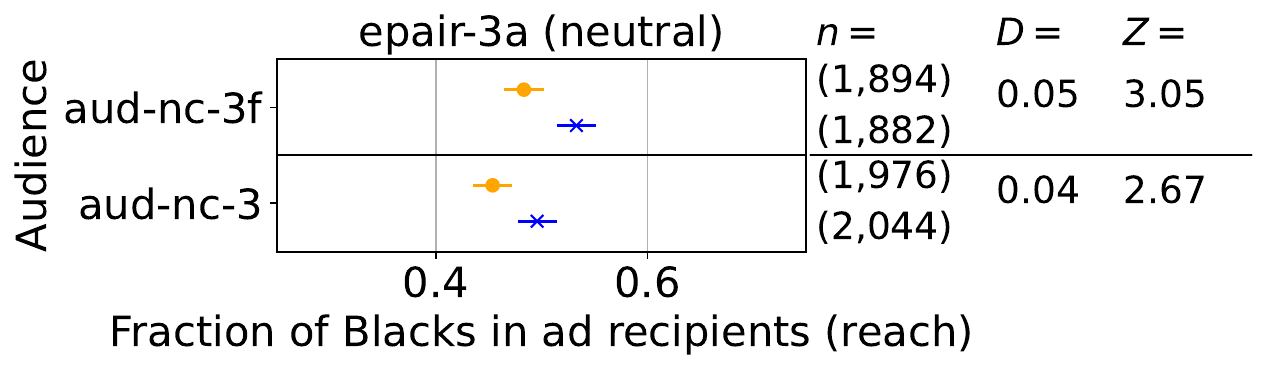}
    \caption{Monroe vs. Arizona State (neutral creatives)}
    \label{fig:epair-3-summary}
  \end{subfigure}
\end{minipage}%
\begin{minipage}{.45\textwidth}
  \centering
  \begin{subfigure}{0.95\columnwidth}
    \includegraphics[width=1\columnwidth]{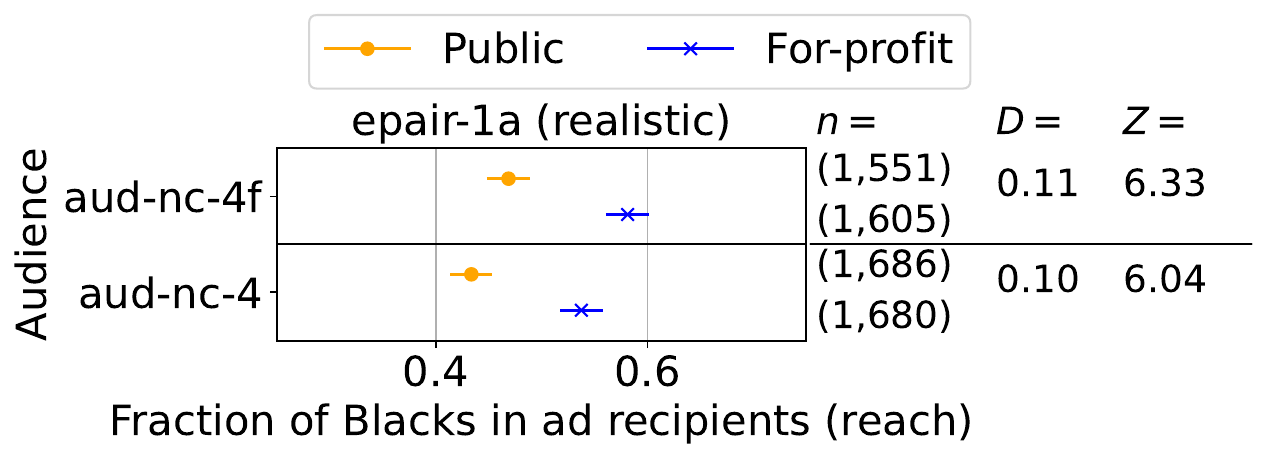}
    \caption{Strayer vs. Colorado State (realistic creatives)}
    \label{fig:epair-1-summary-realistic}
  \end{subfigure}
  \begin{subfigure}{0.95\columnwidth}
    \includegraphics[width=1\columnwidth]{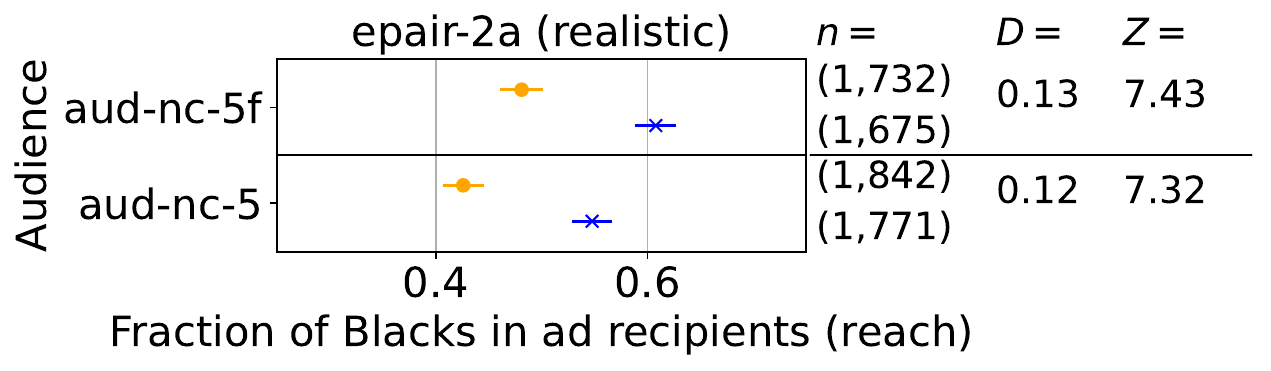}
    \caption{American Inter. vs. Fort Hays (realistic creatives)}
    \label{fig:epair-2-summary-realistic}
  \end{subfigure}
  \begin{subfigure}{0.95\columnwidth}
    \includegraphics[width=1\columnwidth]{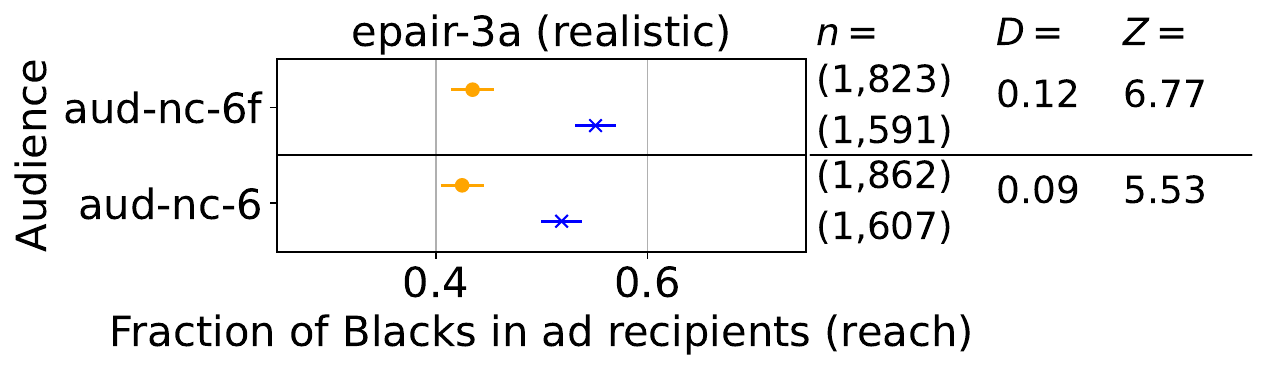}
    \caption{Monroe vs. Arizona State (realistic creatives)}
    \label{fig:epair-3-summary-realistic}
  \end{subfigure}
\end{minipage}
  \caption{Results for Meta's delivery of educations ads for neutral creatives (left) and realistic creatives (right).  Bars show 95\% confidence intervals around each fraction. $n$ is the number of individuals each ad was shown to. $D$ is the difference between fraction of Blacks seeing for-profit and public school ads. $Z$ is the test statistic for significance of this difference.
  An audience named ``{aud-nc-*}'' is built using Black individuals from DMA group 1 (Table~\ref{tab:voter_data_list}) and White individuals from group 2; ``{aud-nc-*f}'' is a flipped version.}
  \label{fig:racial_skew_results}
\end{figure*}

We use $D$ to represent the difference between $s_{f,b}$ and $s_{p,b}$:
  $D = s_{f,b} - s_{p,b}$.
We apply one-sided Z-test for difference in proportions to test whether the difference
  between the two fractions is statistically significant, where our null hypothesis is $D = 0$ and our alternate hypothesis is $D > 0$. The test statistic is given by:

\begin{equation}\label{eq:true_Z_formula} %
Z = \frac{D}{{ {\V{SE} }}} \quad \text{where} \quad {\V{SE}} = \sqrt{\hat{s}_{b} (1 - \hat{s}_{b}) \left(\frac{1}{n_{f}} + \frac{1}{n_{p}}\right)}
\end{equation}

\noindent where $n_{f} = n_{f,b} + n_{f,w}$ and $n_{p} = n_{p,b} + n_{p,w}$ and
$\hat{s}_{b}$ is the fraction of Black users in combined set of all people that saw at least one of the two ads:
 $\hat{s}_{b} = \frac{n_{f,b} + n_{p,b}}{n_{f} + n_{p}}$.
Finally, we pick a level of significance $\alpha$ (typically, $0.05$),
  to determine the corresponding critical value of $Z_\alpha$ from the Z-table
  for standard normal distribution,
  and conclude that there is a statistically significant racial skew
  in the ad delivery algorithm if $Z > Z_\alpha$.
We use a 95\% confidence level ($Z_{\alpha} = 1.64$) for all of our statistical tests.
This hypothesis assumes the samples of individuals that see the ads
  are independent and that $n$ is large.
The delivery audiences may have overlapping samples
  that are dependent if the same person sees both ads, but
  we target a large audience to minimize
  such an outcome.
The sample sizes vary by each experiment,
  but they are at least 1,500
  as shown in the ``$n$'' column in Figure~\ref{fig:racial_skew_results}.

To check robustness of our results,
  we additionally apply a multiple-test correction.
We use Holm's method~\cite{Holm1979},
  a statistical technique that corrects
  for the issue of multiple testing by
  adjusting the threshold for statistical significance
  depending on the number of tests.

\section{Experiments and Results}
   \label{sec:bias_result}

We apply the methodology we developed 
  to create, run and compare the delivery of ads on Meta for the
  pairs of for-profit and public schools.
We ran all ads between April 2023 and April 2024.
Our hypothesis is that,
  if a platform's algorithm for education ad delivery is discriminatory,
  the for-profit school ad will
  be delivered to a disproportionately larger fraction of Black users than the simultaneously run public school ad.
Our findings demonstrate evidence of racial discrimination in Meta's algorithms.  
 We make data from our experiments
  publicly available at~\cite{EducationAdDeliveryDataset}.

We apply our method to Meta because it is a major ad platform
  with billions of users, and has known risks of
  algorithmic bias, as documented in
  prior academic work~\cite{Speicher2018, Ali2019a, Ali2019b, Imana2021},
  civil rights audits~\cite{FbCivilRightsAudit2020},
  and the settlement with the U.S.~Department of Justice~\cite{FacebookvsHUD2}.
  
\subsection{Demonstrating Discriminatory Ad Delivery using Neutral Ad Creatives}
  \label{sec:netural_ads} 

 We present results of our experiments using neutral ad creatives (see Figure~\ref{fig:education_ads_for_neutral}  and Figure~\ref{fig:education_ads_pub_neutral} for examples of such creatives for one pair of for-profit and public schools).
 
Recall that Meta reports location of ad recipients,
  but we are interested in the racial breakdown of the recipients.
As discussed in \autoref{sec:audience_education},
  we repeat each experiment
  on two audiences by flipping the DMAs from which we pick White and Black individuals.
Evaluating both combinations allows us to factor out location as a confounding
  factor.
By applying this procedure to the three pairs of schools
  we identified in~Table~\ref{tab:education_ad_list},
  we run a total of six experiments.
In \autoref{sec:state_proxy_results},
  we present additional experimental results where we use states, instead of DMAs,
  as a way to construct audiences with a location - race correspondence.

We show our direct observations of ad delivery by race for the experiments in
  the left column of Figure~\ref{fig:racial_skew_results}.
Each row of that column shows the result of an experiment of running ads for a pair of schools,
  one public (top, orange) and the second -- for-profit (bottom, blue).  
For each pair of ads, we report $D$ computed 
  by subtracting the fraction of Blacks who saw the public
  school ad from the fraction of Blacks who saw the for-profit school ad
  ($D = s_{f,b} - s_{p,b}$).
We expect $D$ to be positive if there is discriminatory 
  ad delivery consistent with the de-facto racial skew in the
  demographics of the schools' student body (\autoref{tab:education_ad_list}).

\begin{figure*}
  \begin{minipage}{0.69\linewidth}
    \begin{subfigure}{0.45\columnwidth}
\includegraphics[width=\columnwidth]{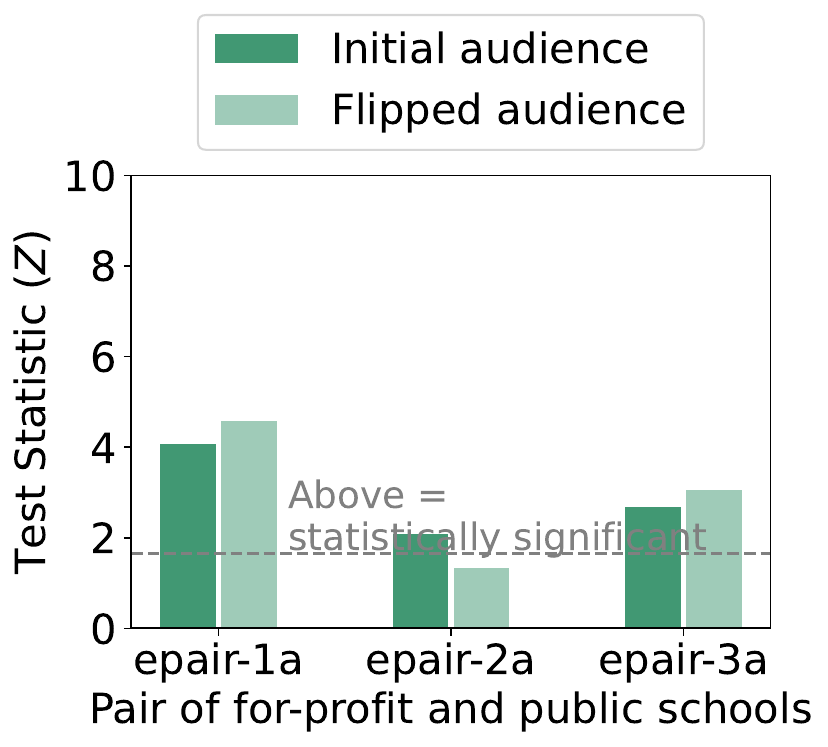}
\caption{Using neutral ad creatives.}
 \label{fig:racial_skew_results_bar}
\end{subfigure}
\begin{subfigure}{0.45\columnwidth}
\includegraphics[width=\columnwidth]{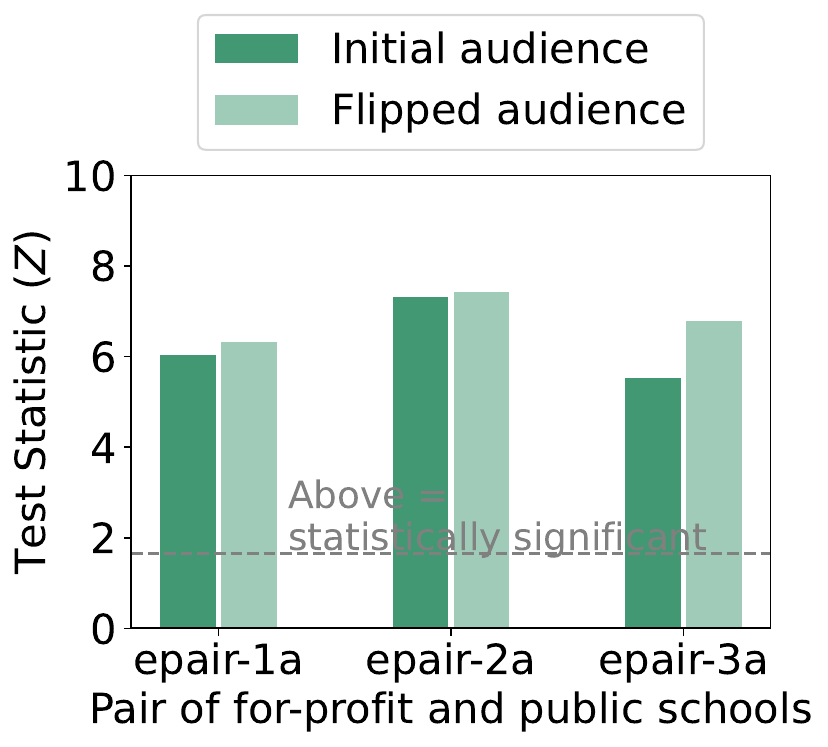}
\caption{Using realistic ad creatives.}
 \label{fig:racial_skew_results_bar_faces}
\end{subfigure}
  \end{minipage}
  \hfill
  \begin{minipage}{0.3\linewidth}
    \caption{Statistical significance of racial skew in delivery of education ads on Meta. The test statistic is computed based on the racial skew measured in~Figure~\ref{fig:racial_skew_results}.  The racial skew in delivery between a pair of ads is statistically significant if the test statistic bar is above the horizontal line (which corresponds to a 95\% confidence level: $Z_{\alpha} = 1.64$). Each bar corresponds to an experiment.}
    \label{fig:racial_skew_results_all}
  \end{minipage}
\end{figure*}

We find that the ads for for-profit school in all six experiments with neutral creatives
  are shown to a higher fraction of Black users ($D$ is positive),
  as shown in Figure~\ref{fig:epair-1-summary}, Figure~\ref{fig:epair-2-summary} and Figure~\ref{fig:epair-3-summary}.
Moreover, the racial skew is statistically significant
  in five
  out of the six experiments.
To illustrate, consider Figure~\ref{fig:racial_skew_results_bar}
  which reports the test statistics for each pair of ads.
We compute the significance test statistics using the
  formula in \autoref{eq:true_Z_formula},
  and we compare it to the threshold of statistical significance,
  shown by the dotted horizontal line.
The racial difference in ad delivery for a pair of ads
  is statistically significant
  for a pair when a test statistic is above the horizontal line.
The test statistic for epair-1a and epair-3a
  crosses the threshold for both the initial and flipped audiences.
For the second pair (epair-2a),
   the skew in ad delivery we observe is not large
  enough to be statistically significant for the flipped audience.
One possible explanation is that the de-facto racial skew
  in the demographic
  of students for the second pair is smaller than the
  other two pairs (see Table~\ref{tab:education_ad_list}),
  hence resulting in a smaller skew in ad delivery.
Our conclusions remain the same after
  applying Holm's correction for multiple hypothesis testing
  over the six tests.  

Our finding of algorithm-induced skew in majority of the cases
  suggests \emph{Meta's ad platform racially discriminates
  in the delivery of education ads}.
Because we measured the relative difference in the delivery of paired ads,
  the bias we observed cannot be explained by confounding
  factors such as targeting choices, differences in who is online,
  competition from concurrently running ads, or market forces.
Such factors affect both the for-profit and public college ads
  in our experiments equally.
Therefore, the racial bias we measured in the delivery of the education ads
  is a product of choices made by Meta's ad delivery algorithms.
  
\subsection{Demonstrating Amplified Skew Using Realistic Ad Creatives}
  \label{sec:realistc_ads}

We next re-run the ads for the same schools as in \autoref{sec:netural_ads}, but using realistic ad creatives 
  (sampled from those actually used by schools),
   to measure by how much
  racial skew in ad delivery increases in the real world.
Our expectation is that creatives that include faces of students
  may propagate assumptions about the racial mix of student body,
  giving a platform's ad delivery algorithm additional, implicit information
  to determine to whom the ad may be relevant.
Prior work has shown this additional factor may \emph{increase} the skew~\cite{Levi2022} and studied the potential for discrimination through selective use of images in the employment advertising context~\cite{Nagaraj2023}.

Using the method described in \autoref{sec:edu_ad_creatives},
  we observe that the for-profit schools have more ads depicting Black faces
  than ads depicting White faces, and vice versa for the public schools.  
We thus pick a representative ad that includes the face of a (perceived) Black person for the for-profit school and a face of a White person for the public school, respectively.
  
We find that the racially skewed delivery of the ads for the for-profit school
  is further amplified for the realistic ad creatives (Figure~\ref{fig:racial_skew_results}; right column).
We can see the amplification by comparing the levels of skew
  ($D$) in the two columns of the figure.
For example,
  for epair-1a,
  the skew is $D=0.07$ when measured using neutral ad creatives (``aud-nc-1f'' in Figure~\ref{fig:epair-1-summary}),
  but is larger, $D=0.11$, when using realistic ad creatives (``aud-nc-4f'' Figure~\ref{fig:epair-1-summary-realistic}).
We illustrate how the amplified skew affects our conclusion in Figure~\ref{fig:racial_skew_results_all}.
For each pair of schools,
  the skew we measure using realistic ad creatives (bars in Figure~\ref{fig:racial_skew_results_bar_faces})  are larger than the corresponding measurement
 using neutral ad creatives (bars in Figure~\ref{fig:racial_skew_results_bar}).
For the realistic ad creatives,
  we observe a statistically significant skew for all three pairs,
  including epair-2a for which the skew we measured using neutral
  ad creatives is not
  statistically significant.
The results for all three pairs remain statistically significant after applying Holm's correction
  to the family of tests. %

This result shows that,
  in addition to perpetuating historical racial biases
  associated with the schools,
  \emph{platforms amplify implicit cues in ad creatives used
  by schools}.
Taken together,
  the advertiser's choice of ad creatives  and
  algorithmic steering that further amplifies those choices point to
  a serious impact of ad delivery algorithms in
  shaping access to education opportunities.

\subsection{Experiments Using Schools with Prior Legal Scrutiny}
  \label{sec:exp_legal_scru}

We next apply our methodology
  to examine whether Meta's algorithm
  racially biases the delivery of ads for schools with prior legal scrutiny
  for their marketing or recruitment practices.
Prior work has suggested
  that opportunities at such
  institutions can provide poor financial outcomes
  to students who enroll in these schools~\cite{Deming2013, Looney2015}.

We identify schools that have previously received legal scrutiny for their marketing and recruitment practices
  from a list published by
  Veterans Education Success in 2018
  for the U.S. Department of Education National Advisory Committee on Institutional Quality and Integrity~\cite{DOESuedSchools}.
From this list,
  we pick three private universities with the largest number of students:
  DeVry, Grand Canyon and Keiser.
DeVry and Grand Canyon are currently
  private for-profit universities;
  Keiser was a private, for-profit university previously
  but changed to non-profit status in 2011.
We then evaluate delivery of ads advertising educational opportunities at those schools using our
  methodology.
Our goal is to test whether Meta's algorithms result in skewed delivery
  even if the schools have changed
  their marketing practices.
  
\begin{table*}
\caption{Racial make-up of private schools with prior legal scrutiny and the public schools we pair them with.}
\label{tab:legal_schools_data}
\centering
\small
 \begin{tabular}{ p{5em}|p{18em}|p{18em} }
Pair ID & Private School & Public School \\ 
\hline 
epair-1b & DeVry University \newline (B=26\% W=44\%, O=29\%) (Admit: 44\%) & Colorado State University \newline (B=7\%, W=64\%, O=29\%) (Admit: 98\%)   \\ %
\hline 
epair-2b & Grand Canyon University\newline  (B=16\% W=48\%, O=37\%) (Admit: 81\%) &  Fort Hays State University \newline  (B=3\%, W=50\%, O=47\%) (Admit: 91\%)  \\  %
\hline 
epair-3b & Keiser University \newline  (B=19\% W=30\%, O=51\%) (Admit: 97\%) &  Arizona State University \newline  (B=7\%, W=58\%, O=35\%) (Admit: 73\%)  \\  %
\end{tabular}
\end{table*}
  
We apply the methodology
  described in~\autoref{sec:education_method},
  but we modify the school-selection criteria
  to consider the three schools we selected as described above.
We continue to match the selected schools with the same three public schools used in our earlier experiments, following the process outlined in \autoref{sec:pair_ads_education}.
Table~\ref{tab:legal_schools_data} provides an overview of the school pairings we used in this experiment.
Similar to \autoref{sec:realistc_ads}, we use realistic ad
  creatives that are taken from each school's Meta ad library page
  (see \autoref{sec:app_ad_creatives} for ad screenshots).
As in our previous experiments,
  we replicate the experiments on a ``flipped'' audience
  to ensure location does not skew our results.
 
\begin{figure}
\centering
\begin{minipage}{0.46\textwidth}
  \centering
   \begin{subfigure}{0.95\columnwidth}
    \includegraphics[width=1\columnwidth]{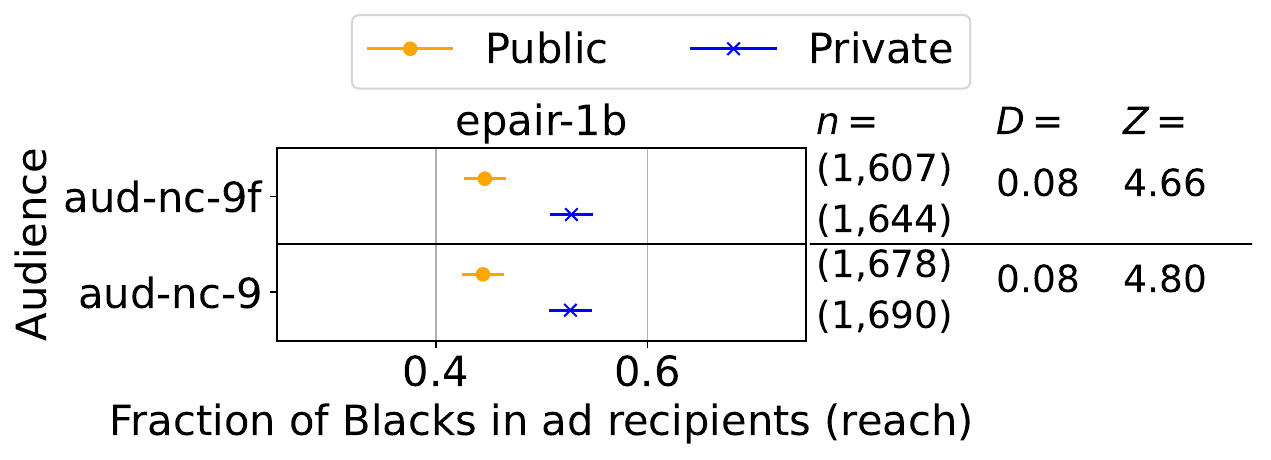}
    \caption{DeVry University vs. Colorado State}
    \label{fig:epair-5-summary}
  \end{subfigure}
  \begin{subfigure}{0.95\columnwidth}
    \includegraphics[width=1\columnwidth]{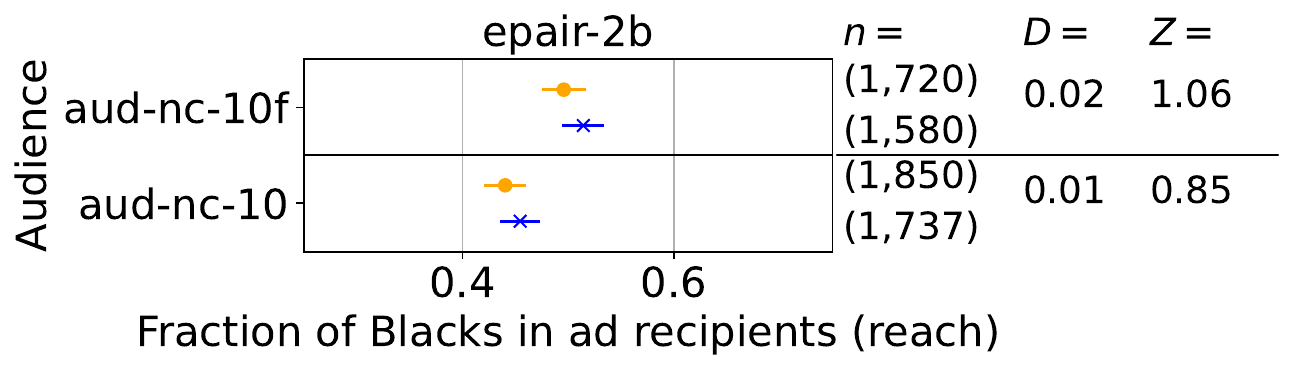}
    \caption{Keiser University vs. ASU}
    \label{fig:epair-6-summary}
  \end{subfigure}
  \begin{subfigure}{0.95\columnwidth}
    \includegraphics[width=1\columnwidth]{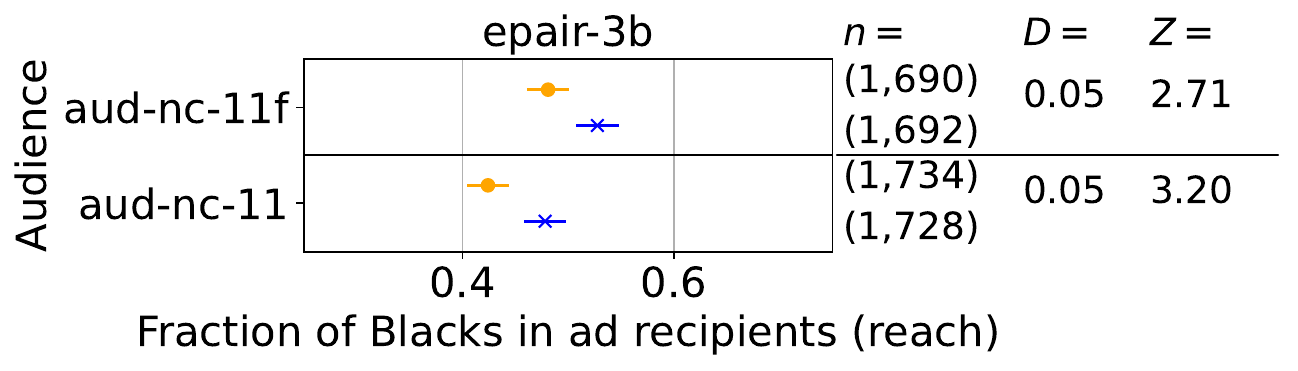}
    \caption{Grand Canyon University vs. Fort Hays}
    \label{fig:epair-7-summary}
  \end{subfigure}
\end{minipage}
\hfill
\begin{minipage}{0.43\textwidth} %
  \centering
    \begin{subfigure}{0.7\textwidth}
        \includegraphics[width=\textwidth]{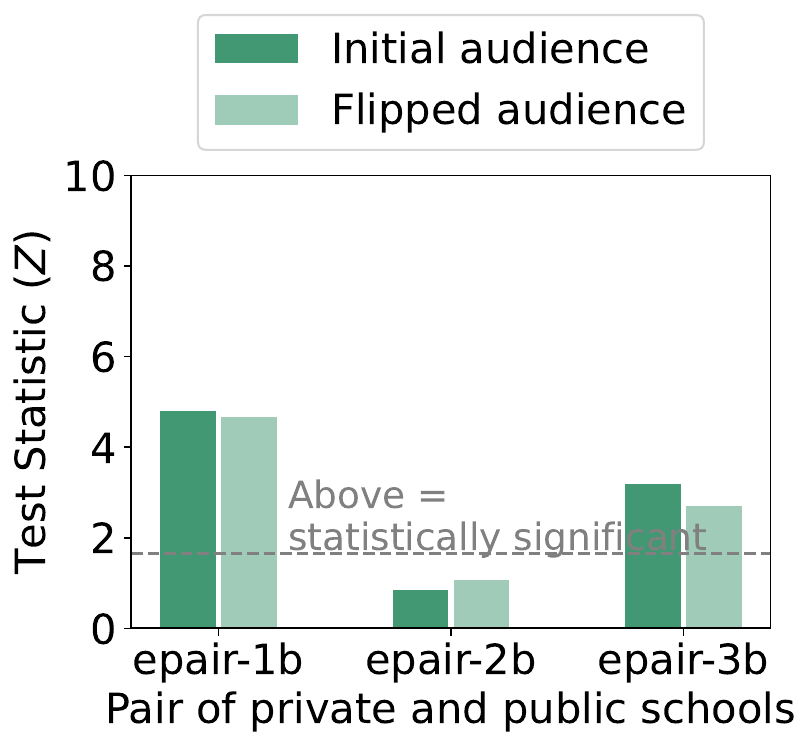} 
        \caption{Statistical significance of skew}
        \label{fig:stat_sign_pred_schools}
    \end{subfigure}
\end{minipage}
 \caption{Skew measured by comparing the delivery of ads for private schools with prior legal scrutiny and public schools.}
  \label{fig:racial_skew_results_legal_mult_aud}
\end{figure}

We find that, for all three pairs,
  the ad for the school with prior legal scrutiny is shown to
  relatively larger fraction of Blacks
  compared to the public school ad
  (Figure~\ref{fig:racial_skew_results_legal_mult_aud}; left column).
For example,
  for the first pair of schools
  (top row of Figure~\ref{fig:epair-5-summary}),
  the public school ad (orange circle)
  is delivered to 44\% Blacks whereas the
  ad for the private school (blue cross mark)
  is delivered to relatively larger fraction of Blacks (52\%).
For two out of the three pairs,
  the racial skew in delivery is statistically significant (see Figure~\ref{fig:stat_sign_pred_schools})
and remains statistically significant after applying Holm's correction.

Our findings show that \emph{Meta's algorithms
  disproportionately deliver educational opportunities
  at the selected schools based on race}.
Our study illustrates how platform-driven discrimination can happen in practice,
  as the schools are active advertisers on Meta’s ad platform.

\section{Discussion}
Our work has implications for
  how platforms shape access to education opportunities,
  the potential legal liability that they may incur as a result,
  and the need for platforms, researchers, and regulators
  to follow a more holistic approach
  to identifying and addressing the issues of bias and discrimination in ad delivery.
   
\subsection{Platforms Shaping Access to Education Opportunities}

Our findings highlight the negative impact of ad delivery algorithms
  in shaping access to education opportunities,
  adding to existing concerns about the disadvantages
  of attending for-profit schools.
Studies have raised concerns that for-profit colleges
  provide poorer outcomes,
  with their students showing 
  higher loan default rate,
  lower earnings and employment
  than comparable students at other post-secondary institutions~\cite{Deming2013, Looney2015, FederalReserve2023}.  
Kahn et al.~have highlighted the risk that
  for-profit schools result in high debt and negative outcomes,
  particularly for Black students~\cite{Kahn2019}.

The platform-induced racial bias we demonstrate
  illustrates another significant factor that decides
  exposure to education opportunities: ad delivery algorithms.
We show Meta's algorithms deliver
  ads for the for-profit schools to relatively more Black individuals
  than ads for public schools.
The racial difference we observe in the delivery is
  not due to the advertiser's targeting choices
  since we select racially balanced audiences.
It is also not due to market effects or difference in platform use by race since our methodology controls for those.  
Therefore, even if an institution with past skewed racial enrollment aims for racially balanced ad targeting,
  Meta's algorithms would recreate historical
  racial skew in who the ad are shown to, and would do so unbeknownst to the advertisers.
Our findings show it is not enough for schools
  to target their ads equitably;
  platforms also need to ensure their ad delivery
  algorithms are not biased by race.

\subsection{Legal Liability for Ad Platforms}
  \label{sec:legal_liability}

In addition to the harm in selectively showing
  potentially lower-quality education opportunities disproportionally to racial minorities, 
  discriminatory delivery of education ads risks legal
  liability for Meta.
  
Educational opportunities have legal protections that prohibit racial discrimination
  and may apply to ad platforms.
The Civil Rights Act of 1974
   prohibits discrimination by race, color, or national origin~\cite{TitleVI}
   for schools that receive federal funding.
D.C.'s Human Rights Act,
  which provides one of the broadest protections
  among the states in the U.S.~\cite{Lerman1978},
  similarly prohibits discrimination by these and other
  attributes such as sex, age and religion for several
  domains including education~\cite{DCHumanRights}.
 
Our results show Meta should be under legal scrutiny
   for the role its algorithms play in the delivery of education ads.
Because we show discrimination in the
   \emph{outcome} of ad delivery,
   the platform may be liable under the \emph{disparate impact} doctrine of discrimination~\cite{Barocas2016, datta2018discrimination}.
Under this doctrine,
  a claim of discrimination can be made if the algorithmic outcome
  differs significantly by a protected demographic attribute,
  regardless of the source of bias for such outcome~\cite{Barocas2016}.
In this case, the burden for justifying the specific source of bias
  shifts to the platform.
Given our findings of discriminatory ad delivery
  after controlling for conceivable confounding factors,
  these results suggest that Meta may need to either justify the concern,
  or address it with modifications to its algorithms.

\section{Ethical Considerations}
  \label{sec:ethics}

We conduct our audits with consideration of the ethical implications
   to both
   individuals engaging with our advertisements
   and to platforms.
First, we build our audiences with voter registration datasets
  available to the public through election offices of U.S. states.
Since this data is already public, it poses minimal new privacy risks.
Furthermore, we use this public information
  only to build our ad audiences;
  we do not interact directly with the users,
  nor do we receive or collect identifiers about individuals who
  see our ads.
We observe and report only aggregate statistics about the results
  of ad performance.
Although we consider these risks minimal,
  this use of voter datasets may require additional considerations in cases where GDPR applies.
Second,
  while we purchase ads,
  our spending on the purchases is tiny and the benefits of learning about potential discrimination 
  in platform algorithms greatly outweighs
  potential cost they impose on the ad recipients.
While our ads for for-profit schools may substitute for better opportunities,
  our ad spending is vanishingly small (a few hundred dollars) 
  relative to the advertising budgets of for-profit colleges
  (where their median marketing cost to recruit a single student exceeds  \$4k~\cite{Gold2019}).
We minimize any overhead our ads place on their viewers --
  they link to real education opportunities,
  and we select audiences of equal sizes by race.
Our approaches do not harm the platform,
  follow the terms of service and
  use only features and APIs Meta makes available to any advertiser.
Our work was reviewed and approved 
  by the Institutional Review Boards at the University of Southern California (review \#UP-20-00132) and
  at Princeton University (record \#14833).

\section{Conclusion}

Our study demonstrates yet another domain
  and demographic group
  for which platforms shape
  access to important life opportunities.
The racial bias we find in education ad delivery
  shows discriminatory delivery extends beyond the current scope of solutions, which
  have been limited to
  housing, employment and credit domains~\cite{FacebookvsHUD2023},
  and raises the broader questions of what other domains
  with legal concerns of discrimination or ethical concerns of bias need equal level of
  attention from platforms and regulators.
A recent Executive Order towards regulating Artificial Intelligence in the U.S.~identifies
  a number of other domains, such as insurance, healthcare, and
  childcare,
  as domains where AI impacts access to opportunities~\cite{Biden2023, Biden2023Memo}.
Prior studies have also demonstrated that
  ad delivery algorithms shape
  public discourse on topics such as politics~\cite{Ali2019b} and climate-change~\cite{Sankaranarayanan2023}.

Given the range of domains with concerns of algorithmic bias,
  we call on platforms to, first,
  conduct impact assessment of their ad delivery algorithms across all domains relevant to civil rights and societally important topics,
  and publicize their findings.
Second,
  platforms should do more to allow for independent external scrutiny of the impacts of their algorithms.
Current platform transparency efforts
  are limited to data about content, not algorithms~\cite{YoutubeResearchProg, FacebookFORT} and do not provide an infrastructure for experimentation.
Thus the scrutiny of algorithmic impacts so far has been
  limited to intricate black-box audits
  custom designed for each new domain
  where a concern of discrimination arises.
In the rare cases when algorithmic scrutiny has been supported by platforms, it was done only through close
  collaboration between the platforms and select researchers~\cite{Guess2023b, Guess2023a, Nyhan2023, Bailon2023}, without guarantees of full independence.
Our work is evidence that the advocacy in the E.U.~and the U.S.~\cite{Pata2022, DsaApproved} for a wider range of public-interest researchers to be able to scrutinize the algorithms is justified, and indeed, essential to make progress.
A platform-supported
  approach that gives researchers a standardized interface to not only public data but
  to the outputs of the algorithms in a privacy-preserving manner
  is a promising approach towards this goal~\cite{Imana2023, Ojewale2024}. 
  
\begin{acks}
This work work was funded in part by a grant from
  The Rose Foundation and National Science Foundation grants
  CNS-1956435, %
  CNS-1916153, %
  CNS-2333448, %
  CNS-1943584, %
  CNS-2344925, %
  CNS-2319409, %
  and CNS-1925737. %
We are grateful to Jonathan Mayer for feedback on the manuscript.
\end{acks}

\subsection*{Errata}

The arXiv v2 version of this paper was updated from arXiv v1
  to clarify language related to experiments in \autoref{sec:exp_legal_scru}.

\bibliographystyle{ACM-Reference-Format}
%
%
%

\appendix

\section{Experiments Using State Rather than DMA when Constructing Audiences}
  \label{sec:state_proxy_results}

We present additional experimental results that use states instead of DMAs when constructing custom audiences with location - race correspondence.
The results are consistent with those in \autoref{sec:realistc_ads} and \autoref{sec:netural_ads}, with slightly more variation, particularly in
  replications on initial and flipped audiences.
We hypothesize these differences may be due to to state-dependent variation in platform usage, but only Meta has the data to confirm.

For these experiments,
we use publicly available voter datasets from two states: Florida (FL) and
   North Carolina (NC)~\cite{FLVoterData, NCVoterData} (see the summary statistics in Table~\ref{tab:voter_data_list_state_proxy}).
We run each of the ads
  on two audiences --
  with Black individuals and White individuals from FL and NC, respectively, and then
  ``flip'' it with White individuals from FL and Black individuals from NC.
A partition named ``{aud1}''
  indicates we included Black individuals from Florida,
  White individuals from North Carolina,
  and a partition named ``{aud1f}'' is a flipped
  version of the audience.

 \begin{table}[h]
\caption{List of voter datasets we use to construct ad audiences using state as a proxy for race.}
\label{tab:voter_data_list_state_proxy}
\centering
\begin{tabular}{ |c|c|c| }
\hline 
State  & \makecell{\# of Blacks} & \makecell{\# of Whites} \\
\hline
Florida (FL) & 2,090,303 & 9,438,537  \\ 
North Carolina (NC) & 1,546,944 & 4,842,453 \\ 
\hline
\end{tabular}
\end{table}

  \begin{figure*}
\centering
\begin{minipage}{.5\textwidth}
  \centering
  \begin{subfigure}{0.95\columnwidth}
    \includegraphics[width=1\columnwidth]{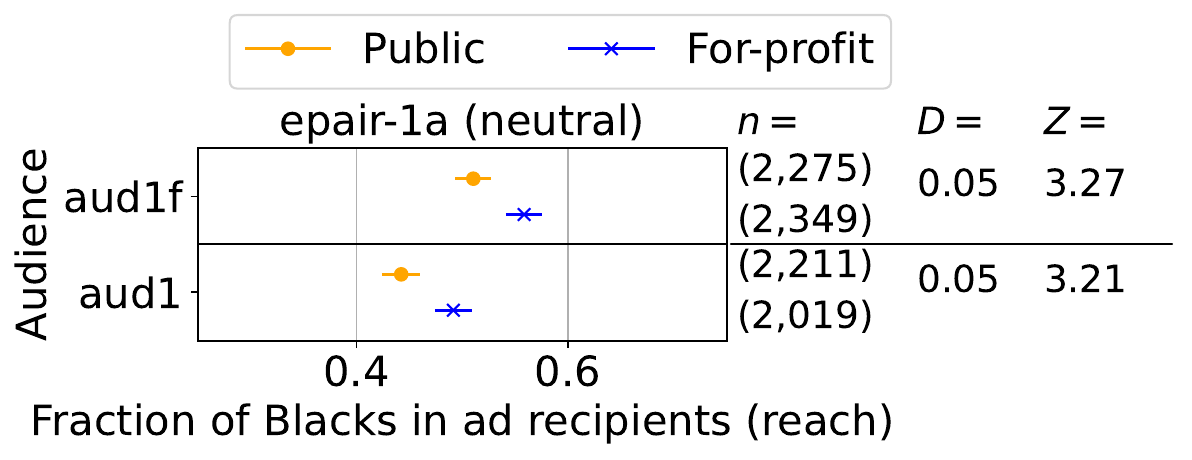}
    \caption{Strayer vs. Colorado State (neutral creatives)}
    \label{fig:epair-1-summary-state}
  \end{subfigure}
  \begin{subfigure}{0.95\columnwidth}
    \includegraphics[width=1\columnwidth]{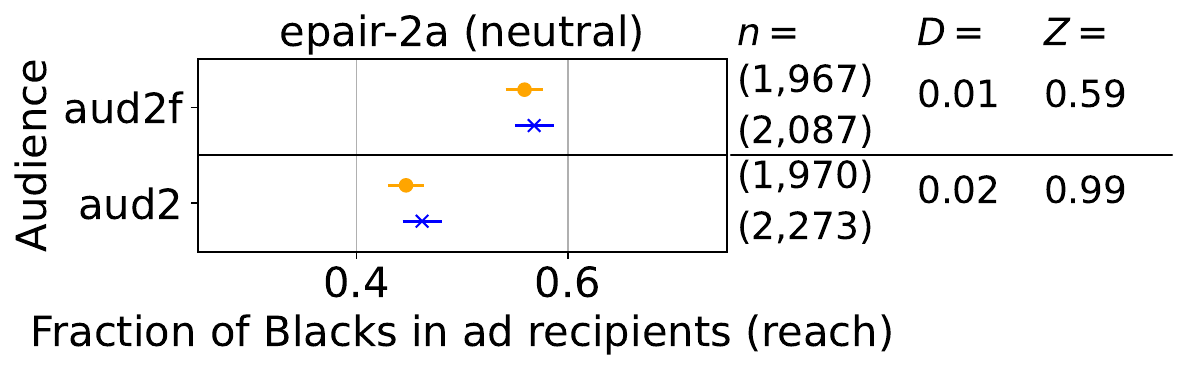}
    \caption{American Inter. vs. Fort Hays (neutral creatives)}
    \label{fig:epair-2-summary-state}
  \end{subfigure}
  \begin{subfigure}{0.95\columnwidth}
    \includegraphics[width=1\columnwidth]{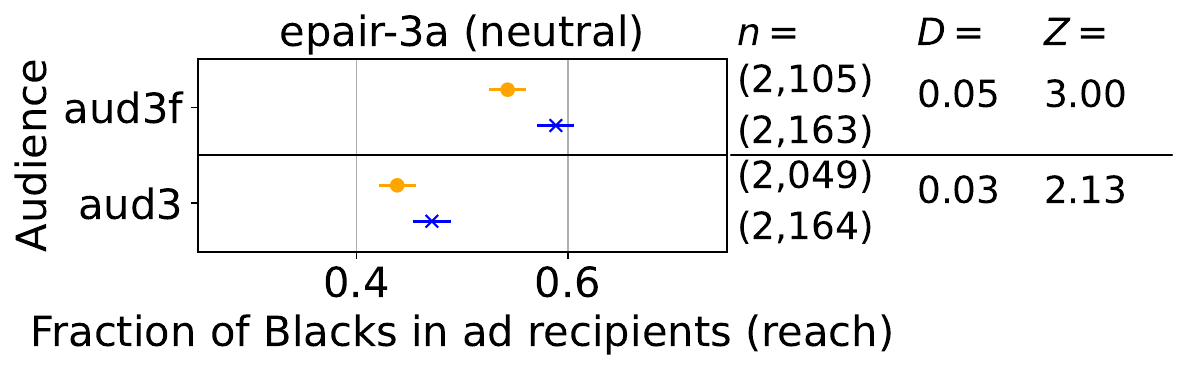}
    \caption{Monroe vs. Arizona State (neutral creatives)}
    \label{fig:epair-3-summary-state}
  \end{subfigure}
\end{minipage}%
\begin{minipage}{.5\textwidth}
  \centering
  \begin{subfigure}{0.95\columnwidth}
    \includegraphics[width=1\columnwidth]{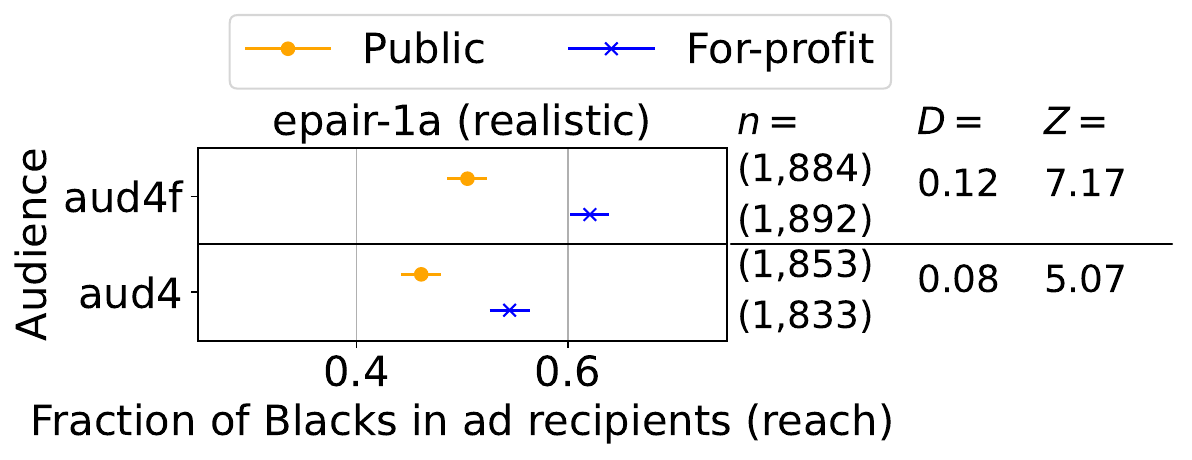}
    \caption{Strayer vs. Colorado State (realistic creatives)}
    \label{fig:epair-1-summary-realistic-state}
  \end{subfigure}
  \begin{subfigure}{0.95\columnwidth}
    \includegraphics[width=1\columnwidth]{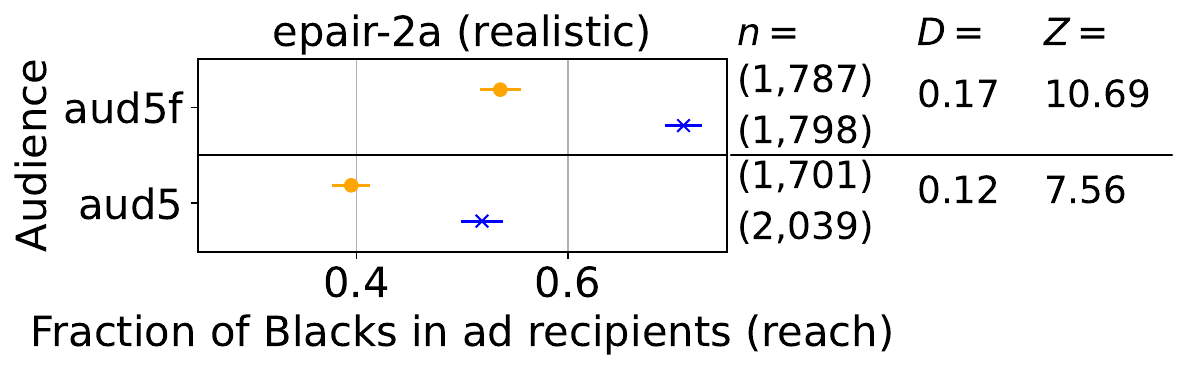}
    \caption{American Inter. vs. Fort Hays (realistic creatives)}
    \label{fig:epair-2-summary-realistic-state}
  \end{subfigure}
  \begin{subfigure}{0.95\columnwidth}
    \includegraphics[width=1\columnwidth]{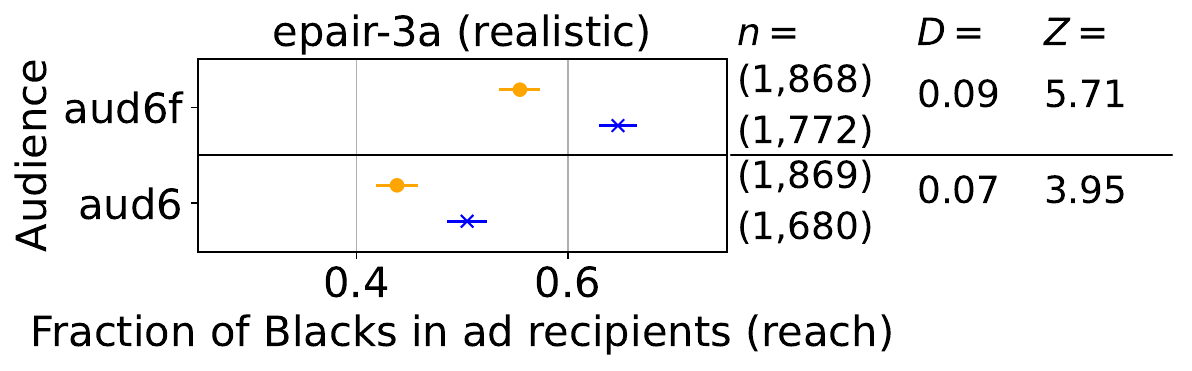}
    \caption{Monroe vs. Arizona State (realistic creatives)}
    \label{fig:epair-3-summary-realistic-state}
  \end{subfigure}
\end{minipage}
  \caption{Additional results for racial skew in the delivery of education ads on Meta. These experiments were conducted using states (NC and FL) as a proxy, hence showing some variation between the flipped and non-flipped results.}
  \label{fig:racial_skew_state_proxy_results}
\end{figure*}

Similar to our findings in \autoref{sec:realistc_ads} and \autoref{sec:netural_ads},
  these experiments show evidence of racial bias in the delivery
  of education ads (left column of Figure~\ref{fig:racial_skew_state_proxy_results}),
  and show the degree of bias in delivery increases when we use realistic
  ad creatives (right column of Figure~\ref{fig:racial_skew_state_proxy_results}).
However, we see some variation in the experiments we replicate
   on flipped audiences for which we expected the results
   to be similar.
For example, in Figure~\ref{fig:epair-2-summary-realistic-state},
  approximately 52\% of recipients of the for-profit school ad are Black
  for aud5, whereas 70\% or recipients are Black when the audience
  is flipped (aud5f).
Because we do not look at absolute numbers but rather relative
  differences in the delivery of the for-profit and public schools,
  these variations do not affects our evaluation of skew in the algorithm.
However,
  the results show using states when constructing audiences can introduce additional
  variations one must keep into account.

\section{Realistic Ad Creatives Used for Private Schools with Prior Legal Scrutiny}
 \label{sec:app_ad_creatives}

\begin{figure*}[h]
\centering
\begin{subfigure}{0.33\linewidth}
\centering
\includegraphics[width=\textwidth]{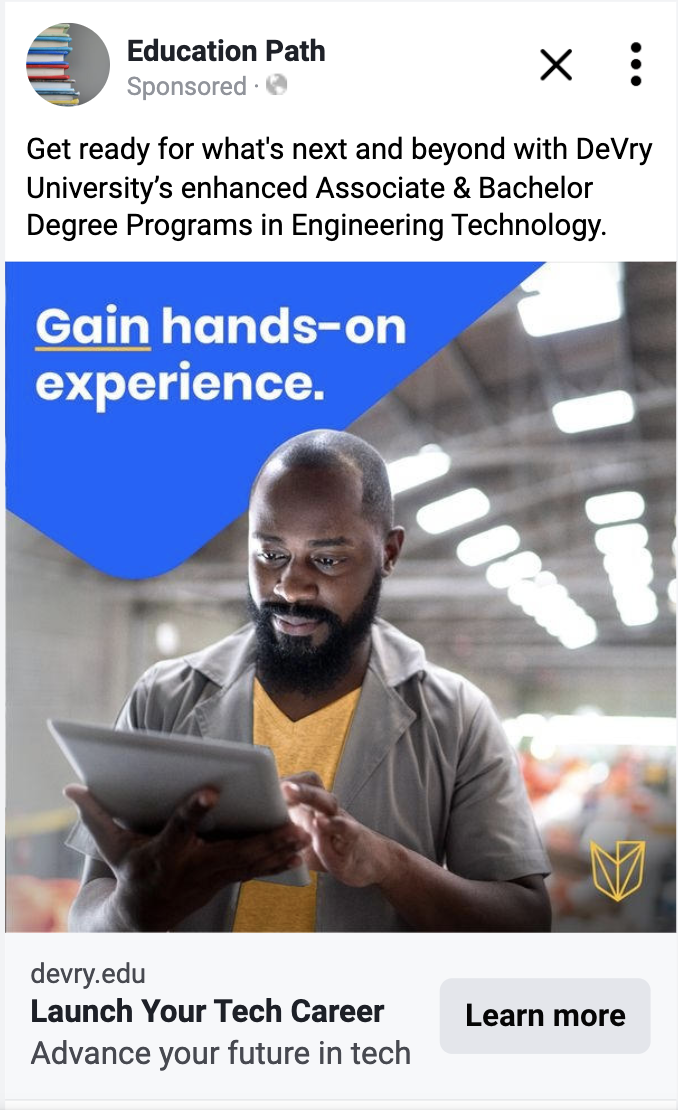}
\caption{DeVry University.}
\end{subfigure}
\begin{subfigure}{0.33\linewidth}
\centering
\includegraphics[width=\textwidth]{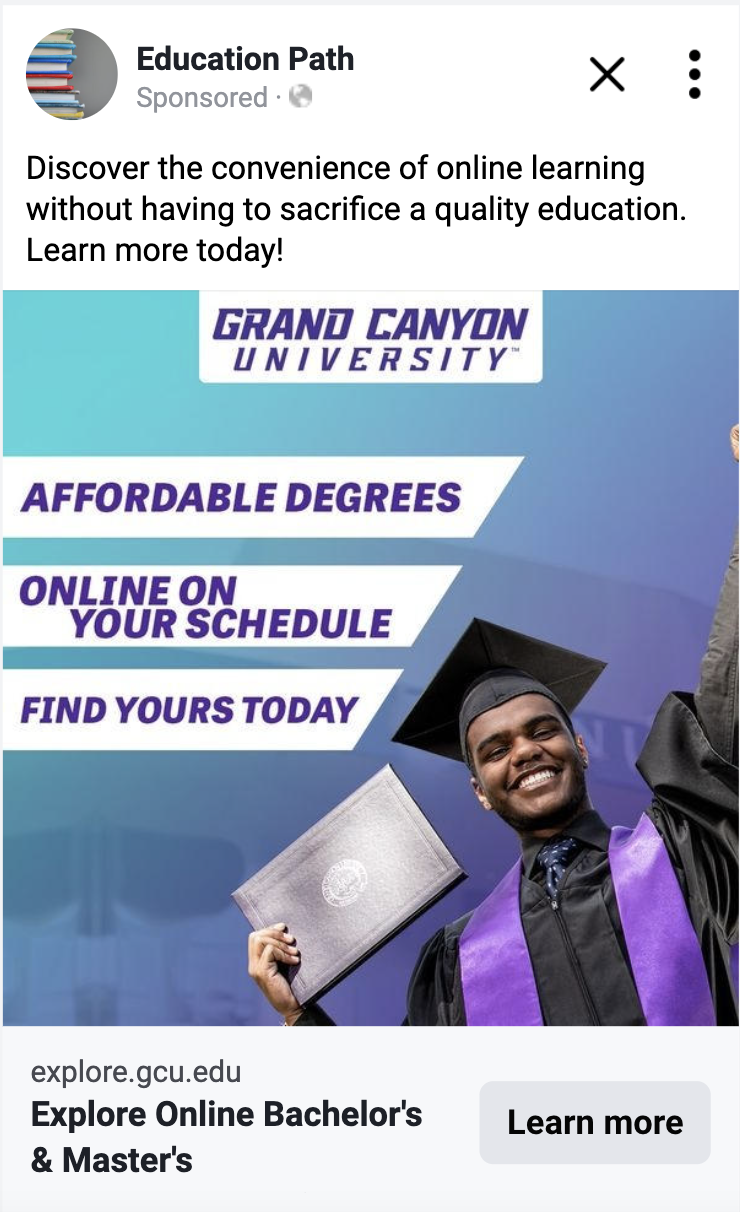}
\caption{Grand Canyon University.}
\end{subfigure}
\begin{subfigure}{0.33\linewidth}
\centering
\includegraphics[width=\textwidth]{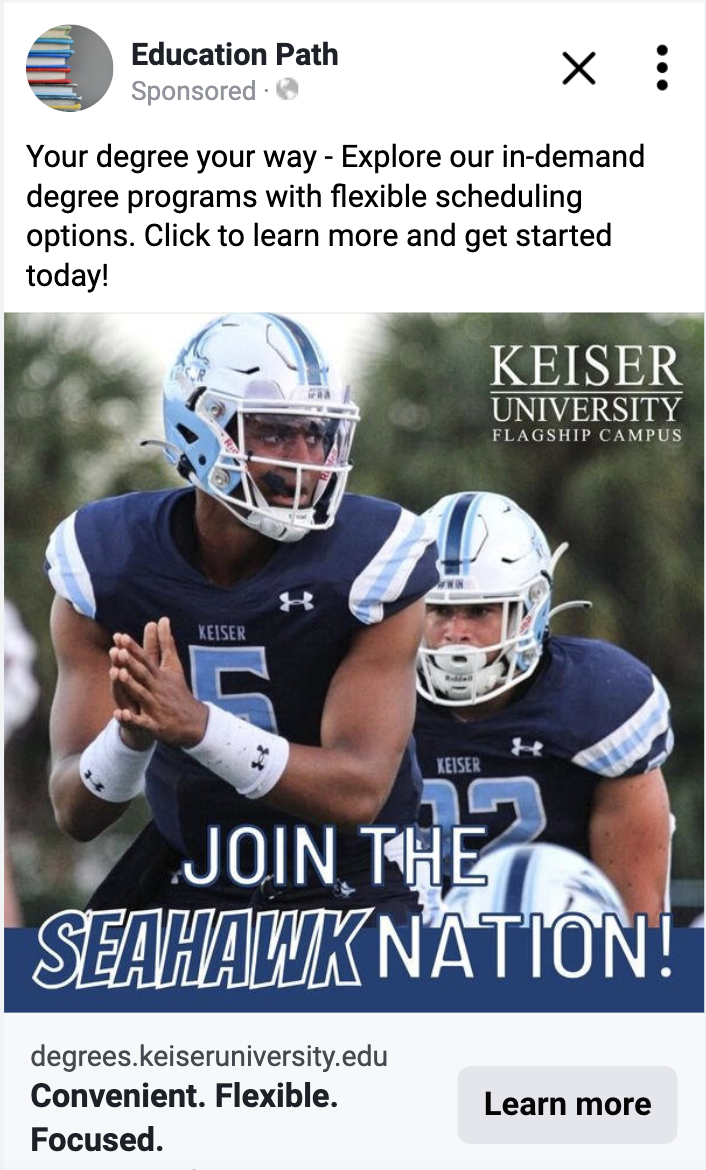}
\caption{Keiser University.}
\end{subfigure}
\caption{Realistic ad creatives used for private schools selected in~\autoref{sec:exp_legal_scru}. The ad texts and images are taken from each school's page on Meta's ad library.}
\label{fig:edu_ad_screenshots3}
\end{figure*}

In \autoref{fig:edu_ad_screenshots3}, we show the realistic ad creatives used for experiments involving
   the private schools selected in \autoref{sec:exp_legal_scru}. These ad creatives are taken from
  each school's page on Meta's public ad library (see \autoref{sec:edu_ad_creatives} for details on 
  creative choice).

\end{document}